\documentclass{aa}  
\usepackage{natbib}
\usepackage{comment}
\usepackage{mathtools}
\usepackage{multirow}
\usepackage{comment}
\usepackage{booktabs}
\usepackage{graphicx}
\usepackage{menukeys}
\usepackage{algorithm}
\usepackage{ulem}
\usepackage[noend]{algpseudocode}
\graphicspath{{Fig/}}
\include{definitions}
\usepackage[colorlinks=true, allcolors=blue]{hyperref}
\usepackage{txfonts}

\newcommand{\arhp}{ArH$^{+}$ }
\newcommand{\ohp}{OH$^{+}$ }
\newcommand{\ohtop}{o-H$_{2}$O$^{+}$ }
\newcommand{\phtop}{p-H$_{2}$O$^{+}$ } 

\usepackage{cuted}
\makeatletter
\renewcommand*\aa@pageof{, page \thepage{} of \pageref*{LastPage}}
\makeatother
\begin{document} 

 \title{ArH$^+$ and H$_2$O$^+$ absorption towards luminous 
 galaxies}

   \author{Arshia M. Jacob$^{1, 2}$
          \and
          Karl M. Menten$^1$
          \and
          Friedrich Wyrowski$^1$
          \and 
          Benjamin Winkel$^1$
          \and 
          David A. Neufeld$^2$
          \and 
          B\"{a}rbel S. Koribalski$^3$
          }

   \institute{Max-Planck-Institut f\"{u}r Radioastronomie, Auf dem H\"{u}gel 69, 53121 Bonn, Germany
   \and 
   Department of Physics and Astronomy, Johns Hopkins University, 3400 North Charles Street, Baltimore, MD 21218, USA 
   \and 
   Australia Telescope National Facility, CSIRO Astronomy and Space Science, P.O. Box 76, Epping, NSW 1710, Australia \\
   \email{ajacob@mpifr-bonn.mpg.de}}

   \date{Received October 28, 2021; accepted December 28, 2021}
  \titlerunning{ArH$^+$ and H$_2$O$^+$ absorption towards luminous nearby galaxies}
   \authorrunning{A. Jacob et al.}
 
  \abstract{Along several sight lines within the Milky Way ArH$^+$ has been ubiquitously detected with only one detection in extragalactic environments, namely along two sight lines in the redshift $z=0.89$ absorber towards the lensed blazar PKS 1830-211. Being formed in predominantly atomic gas by reactions between Ar$^+$, which were initially ionised by cosmic rays and molecular hydrogen, ArH$^{+}$ has been shown to be an excellent tracer of atomic gas as well as the impinging cosmic-ray ionisation rates.} {In this work, we attempt to extend the observations of ArH$^+$ in extragalactic sources to examine its use as a tracer of the atomic interstellar medium (ISM) in these galaxies.}{We report the detection of ArH$^+$ towards two luminous nearby galaxies, NGC~253 and NGC~4945, and the non-detection towards Arp~220 observed using the SEPIA660 receiver on the APEX 12~m telescope. In addition, the two sidebands of this receiver allowed us to observe the $N_{K_{a}K_{c}} = 1_{1,0}-1_{0,1}$ transitions of another atomic gas tracer p-H$_{2}$O$^{+}$ at 607.227~GHz with the ArH$^+$ line, simultaneously. We modelled the optically thin spectra of both species and compared their observed line profiles with that of other well-known atomic gas tracers such as OH$^+$ and o-H$_2$O$^+$ and diffuse and dense molecular gas tracers HF and CO, respectively.} {Assuming that the observed absorption from the ArH$^+$, OH$^+$, and H$_{2}$O$^+$ molecules are affected by the same flux of cosmic rays, we investigate the properties of the different cloud layers. Based on a steady-state analysis of the chemistry of these three species and using statistical equilibrium calculations, we estimate the molecular fraction traced by ArH$^+$ to be ${\sim\!10^{-3}}$ and find that ArH$^+$ resides in gas volumes with low electron densities. We further study the ortho-to-para ratio of H$_{2}$O$^+$ and find that the derived ratios do not significantly deviate from the equilibrium value of three with spin temperatures greater than 15 and 24\,K.}{}

   \maketitle
   
%

\section{Introduction} \label{sec:intro}
Over the duration of its operation, from 2009--2013, the Herschel Space Observatory (HSO) enabled observations of the fundamental rotational transitions of a variety of molecular hydrides and hydride ions, several of them being newly discovered. Many of these transitions cannot be observed from the ground at all because these high frequency lines lie in parts of the sub-millimetre (sub-mm) and far-infrared (FIR) wavelength range that are blocked by absorption in the Earth's atmosphere. One of the many highlights of the Herschel mission, and a real surprise, has been the fortuitous detection of the ${J=1-0}$ and ${J=2-1}$ transitions of argonium, \arhp, in emission towards the Crab Nebula by \citet{barlow2013detection}. Following it, \citet{schilke2014ubiquitous} were able to successfully assign, previously unidentified absorption features near 617\,GHz to ArH$^+$, along the lines-of-sight (LOS) towards five star-forming regions. It turned out that, fortunately, the ${J = 1-0}$ transition of \arhp lies at a wavelength that is accessible with ground-based telescopes at high mountain sites under exceptional weather conditions, and \citet{Jacob2020Arhp} were recently able to detect \arhp towards seven more sight lines in the inner Galaxy using the Atacama Pathfinder Experiment (APEX) 12\,m sub-mm telescope. All of these observations confirm, as first discussed by \citet{schilke2014ubiquitous}, that the \arhp molecular ion exclusively probes diffuse atomic material and that it is ubiquitously present in the Milky Way. This raises questions on the existence and nature of \arhp in extragalactic sources.  

Towards external galaxies, both the ArH$^+$ ${J=1-0}$ and ${J=2-1}$ lines have remained undetected in single side band observations covering their corresponding line frequencies, which were carried out using the Herschel Spectral and Photometric Imaging REceiver \citep[SPIRE,][]{griffin2010herschel} on board the HSO. The apparent non-detection of these lines is likely caused by a combination of effects including smearing by the spectrometer which results in unresolved spectral line profiles alongside effects of blending from nearby lines (such as HCN-v2 (7-6) at 623.3635\,GHz and H$_2$O $2_{2,0}$--$2_{1,1}$ at 1228.79\,GHz). In addition, the ringing noise introduced by uncertainties in the fitting and subtraction of strong lines in the SPIRE FTS spectra using sinc functions can affect the line profiles of the underlying weak absorption. It is for these reasons that ArH$^+$ has remained undetected in observations of extragalactic sources other than PKS 1830$-$211.  
Therefore, very little is known about the nature and abundance of ArH$^+$ outside of the Milky Way. To date, there exists only a single detection of \arhp in extragalactic space, which was carried out by \citet{Mueller2015}. Using the Atacama Large Millimetre/sub-millimetre Array \citep[ALMA,][]{wootten2009atacama}, these authors were able to detect the $J=1-0$ transitions of $^{36}$ArH$^+$ and $^{38}$ArH$^+$ through the intermediate redshift ${z = 0.8858}$ foreground galaxy absorbing the continuum of the gravitational lens-magnified blazar, PKS 1830$-$211 along two different sight lines.

Primarily residing in atomic gas with molecular hydrogen fractions, $f_{\text{H}_{2}}$, between $10^{-2}$ and $10^{-4}$ \citep{schilke2014ubiquitous, neufeld2016chemistry, Jacob2020Arhp}, the abundance of \arhp is sensitive to the X-ray and cosmic-ray fluxes that permeate the surrounding media as its formation is initiated by the reaction between H$_{2}$ and atomic argon ionised by either X-rays and/or cosmic-ray particles. Therefore, observations of the ground state transitions of \arhp provide a unique tool for probing atomic gas and estimating ionisation rates. Regions permeated by a high flux of cosmic-rays can be heated by them to high gas temperatures, which in turn can strongly influence the initial conditions of star-formation and the initial mass function \citep[IMF,][]{papadopoulos2011extreme}. 

In this paper we present our search for \arhp towards three luminous galaxies, Arp~220, NGC~253 and NGC~4945, using the SEPIA660 receiver on the APEX 12\,m telescope. Both systems have been extensively studied over a wide range of wavelengths. In particular, a plethora of molecules have been observed towards these sources, including common hydrides and their cations, for example, OH, OH$^+$, H$_{2}$O, H$_{2}$O$^+$ \citep{Gonzalez2013, van2016ionization, Gonzales2018}. Arp~220 is the archetypical ultra-luminous infrared galaxy (ULIRG). A merging system, it hosts two compact nuclei \citep{baan1995nuclear, rodriguez2005vla} that are surrounded by an immense amount of gas and dust \citep{scoville1997arcsecond, engel2011arp} with dust temperatures between 90 and 160\,K \citep{sakamoto2008submillimeter} and a luminosity of 0.2--$1\times10^{12}\,L_{\odot}$. Notably, the intense starburst activity within the dense interstellar medium (ISM) of its nuclear regions causes stars to form at a rate of up to 50--100\,$M_{\odot}$\,yr$^{-1}$ \citep{Smith1998}, which is ${>\!50}$ times that in the disk of the Milky Way galaxy today \citep{Robitaille2010} and ${\sim\!1000}$ times that in its central molecular zone \citep{Immer2012}. 
NGC~253 is a barred prototypical starburst galaxy part of the Sculptor group, with an infrared luminosity of $1.7\times10^{10}\,L_{\odot}$ \citep{radovich2001far}. Its strong nuclear starburst, drives a ${\sim\!100}\,$pc-scale molecular gas outflow/wind as seen, for example in observations of CO \citep{Bolato2013}. It has been suggested that at the centre of this barred spiral a weak AGN coexists with the strong starburst, an issue that is still under debate \citep[for example see,][]{muller2010stellar, Gutierrez2020}. NGC~4945 is an infrared-bright galaxy with a Seyfert nucleus, signifying an accreting supermassive black hole, in the Centaurus group with a luminosity of $2.4\times10^{10}\,L_{\odot}$ \citep{brock1988far}. It is the brightest Seyfert 2 galaxy and hosts a deeply enshrouded AGN at its centre which is revealed by X-ray emission in the 100-keV sky \citep{Iwasawa1993}. The AGN is surrounded by a strongly absorbing, inclined circumnuclear starburst ring with a radius of ${\sim\! 50}\,$pc \citep{Chou2007}. With comparable star-formation rates of a few times $M_{\odot}$yr$^{-1}$ \citep{Bolato2013, Bendo2016} the similarities (or lack thereof) in the abundances and gas properties traced by ArH$^+$ in these sources will shed light on their nuclear environments.

The observations are described in Sect.~\ref{sec:observations}, followed by a qualitative and quantitative analysis of the data and a discussion of the results in Sects.~\ref{sec:results} and \ref{sec:discussion}. Finally, in Sect.~\ref{sec:conclusions} we discuss our main findings and summarise our results. 

\section{Observations} \label{sec:observations}
Using the Swedish-ESO PI (SEPIA660) receiver \citep{belitsky2018sepia, hesper2018deployable} of the APEX 12\,m sub-mm telescope, we were able to carry out observations of the ${J = 1-0}$ transition of $^{36}$\arhp (hereafter ArH$^{+}$) between 2019 July and August (Project Id: M9519C$\_$103). The SEPIA660 receiver is a two sideband (2SB), dual polarisation receiver that covers a bandwidth of 8\,GHz, per sideband, with a sideband rejection level $>$15\,dB. The observations were carried out in wobbler switching mode, using a secondary wobbler throw of 180$^{\prime\prime}$ in azimuth at a rate of 1.5\,Hz. Combined with the atmospheric stability at the high APEX site, this observing method allows reliable recovery of the sources' continuum levels.  
Properties of our source sample are summarised in Table~\ref{tab:source_properties}.
 
 \begin{table*}
 \begin{center}
 \caption{Properties of studied sources.}
     \begin{tabular}{lrr ccc}
     \hline\hline
          Source & \multicolumn{2}{c}{Coordinates (J2000)} & \multicolumn{1}{c}{D} & \multicolumn{1}{c}{$\upsilon_{\rm helio}$} & \multicolumn{1}{c}{$T_\text{c}$\tablefootmark{a} }\\
          
          & $\alpha$~[hh:mm:ss] & $\delta$~[dd:mm:ss] 
          & [Mpc] & [km~s$^{-1}$] &  [K] \\
          \hline
       Arp~220 & 15:34:57.20 & $+$23:30:11.00 
        & 72.0 &  5434.0 &  0.05 \\
        NGC~253 &  00:47:32.98 & $-$25:17:15.90 & 
        \phantom{0}3.0& \phantom{0}243.0 &0.21\\
        NGC~4945 & 13:05:27.48 & $-$49:28:05.60 & 
        \phantom{0}3.8 &   \phantom{0}585.0 & 0.54 \\
        \hline   
     \end{tabular}
      \label{tab:source_properties}
     \end{center}
     \tablefoot{\tablefoottext{a}{Main-beam brightness temperature of the continuum at 617\,GHz as measured using the SEPIA660 receiver with a HPBW of $10\rlap{.}^{\prime\prime}$3 and a Jy-to-K conversion factor of 70$\pm$6.} }
     \tablebib{Distances are taken from the NASA Extragalactic Database (NED) at \url{https://ned.ipac.caltech.edu}.}
 \end{table*}

We tuned the upper sideband (USB) to a frequency of 618.5\,GHz to cover the \arhp ${J=1-0}$ transition at a rest frequency of 617.525\,GHz. This allowed us to simultaneously observe the $N_{K_{a}K_{c}} = 1_{10}-1_{01}, J = 3/2 - 1/2$, and ${J = 3/2 - 3/2}$ transitions of \phtop at 604.678 and 607.227\,GHz in the lower sideband (LSB), centred at a frequency of 606.5\,GHz. The USB also covers an atmospheric absorption feature close to 620\,GHz, however falls just short of covering the corresponding ${J=1-0}$ transition of $^{38}$ArH$^+$ at 616.648\,GHz. Our observations were carried out under excellent weather conditions, with precipitable water vapour (PWV) levels between 0.25 and 0.41\,mm, corresponding to an atmospheric transmission better than or comparable to 0.5 in both sidebands and a mean system temperature of 1874\,K, at 617\,GHz. On average we spent a total (on+off) observing time of 4.6\,hours towards each source. The typical half power beam-width (HPBW) is $10\rlap{.}^{\prime\prime}$3 at 617\,GHz. The HPBW at this frequency corresponds 0.15\,kpc and 0.19\,kpc at the distances towards NGC~253 and NGC~4945, respectively. Therefore, the beam probes only the nucleus and foreground disk gas of these two galaxies. We converted the spectra into main-beam brightness temperature scales by using an antenna forward efficiency, $F_{\text{eff}}$, of 0.95 and a main-beam efficiency, $B_{\text{eff}}$, of 0.41 (determined from observations of Mars). As a backend we used an evolved version of the MPIfR built Fast Fourier Transform Spectrometer \citep[XFFTS,][]{Klein2012} which provided, over the entire 2$\times$8\,GHz bandwidth, a generic spectral resolution of 61\,kHz (corresponding to 30\,m\,s$^{-1}$ at 617\,GHz). The calibrated spectra, smoothed to a velocity resolution appropriate for our sources (${\Delta\upsilon \sim\!4.5}\,$km~s$^{-1}$), were subsequently processed using the GILDAS/CLASS software\footnote{Software package developed by IRAM, see \url{https://www.iram.fr/IRAMFR/GILDAS/} for more information regarding GILDAS packages.} and first order polynomial baselines were removed. 
Our search for \arhp and \phtop towards Arp~220 was unsuccessful and we do not detect any lines in the sideband down to a noise level of 5\,mK at a spectral resolution of 4.5\,km~s$^{-1}$. Therefore, we do not include Arp~220 in our analysis but quote a 3$\sigma$ upper limit of 15\,mK for the \arhp ${J=1-0}$ and both \phtop transitions in this source. The non-detection of ArH$^+$ towards Arp~220 might be because this system's X-ray emission is not capable of ionising Ar or a combined effect of sensitivity and the low continuum level.

In addition to the results of the \arhp and \phtop observations newly presented in this work, we use complementary archival HIFI/Herschel data of other well known tracers of atomic gas, namely, \ohp and o-H$_{2}$O$^+$, which were acquired as a part of the Herschel EXtraGALactic (HEXGAL) guaranteed time key project (PI: R. G\"{u}sten), published by \citet{van2016ionization} and also of the diffuse molecular gas tracer HF observed by \citet{monje2014hydrogen}. 
The HIFI HPBWs at the frequencies of the OH$^+$, o-H$_{2}$O$^+$, and HF transitions studied in the above works are comparable to one another at 22$^{\prime\prime}$, 20$^{\prime\prime}$, and 17$^{\prime\prime}$, respectively.

The spectrum of the $N,J\!=\!1,2\!-\!0,1$ OH$^+$ transitions near 971~GHz towards NGC~253, observed under the HEXGAL project (presented by \citealp{van2016ionization}) is saturated over almost the entire range of heliocentric velocities associated with the central parts of the galaxy, between 185 and 235\,km~s$^{-1}$. This makes the subsequent determination of OH$^+$ column densities from the observed spectrum extremely difficult. Therefore, in this work, towards NGC~253 we instead use observations of the $N, J\!=\!1,1-0,1$ transitions of OH$^+$ near 1033~GHz observed in 2010 July using the now decommissioned dual-channel 1.05\,THz receiver on the APEX 12~m telescope \citep{Leinz2010}. The observations were carried out under excellent weather conditions with PWV between 0.15 and 0.25\,mm. The fast Fourier Transform spectrometer \citep[FFTS,][]{Klein2006} provided a spectral resolution of ${\sim\!0.053}$\,km~s$^{-1}$ (183\,kHz) over a 2.4\,GHz bandwidth for the 1.05\,THz channel which was later smoothed to a velocity resolution of 4.5\,km~s$^{-1}$. The spectra were calibrated on main-beam temperature scale using a main-beam efficiency of ${(\sim\!21\pm7)\%}$ as determined from observations of Uranus. The observed spectrum shows a narrow absorption feature at 320~km~s$^{-1}$ which likely arises from blending with contaminant species along the sight line. In order to confirm the association of this feature to a species other than OH$^+$ we cross-checked the 1033~GHz OH$^+$ spectrum along the LOS towards the extensively studied Galactic source, Sgr~B2~(M), observed under the HEXOS Herschel guaranteed time key project \citep{bergin2010herschel}. From this comparison we assign this absorption feature near 1032.783~GHz as being likely associated with the $^{13}$CH$_2$CHCN (21$_{10,11}$--20$_{9,12}$) transition. In addition we also observe a weaker absorption feature near 12~km~s$^{-1}$ which potentially arises from the high-lying ($37_{9,29}$--$37_{6,32}$) transition of C$_2$H$_5$CN at 1033.868~GHz. Furthermore the high excitation lines of similar complex organic molecules (COMs) found in ALMA Band 6 and 7 observations towards NGC~253 by \citet{Mangum2019} makes it likely that the OH$^+$ spectrum at 1033~GHz is contaminated by the high-lying transitions of these species. 

We also compare the line profiles of the \arhp line with those of the H{\small I} 21\,cm line and subsequently determine the \arhp abundances. For this we use archival interferometric data of H{\small I} absorption and emission for NGC~253 and NGC~4945, observed using the Australia Telescope Compact Array (ATCA) with beam sizes of $4\rlap{.}^{\prime\prime}9\times10\rlap{.}^{\prime\prime}3$ and $7\rlap{.}^{\prime\prime}9\times9\rlap{.}^{\prime\prime}4$, respectively. The H{\small I} column densities were determined as described in \citet{winkel2017hydrogen}, by combining the absorption profiles with emission line data. The results of this H{\small I} analysis, resulting in determinations of the optical depth, spin temperatures, and H{\small I} column densities along with the corresponding H{\small I} emission and absorption spectra is given in Appendix~\ref{appendix:hi_analysis}. In addition we also present CO $J=(1-0)$ emission spectra towards NGC~253 and NGC~4945, previously published in \citet{Houghton1997} and \citet{Curran2001}, respectively, for comparison\footnote{The spectra were extracted using the WebPlotDigitizer tool available on \href{http://arohatgi.
info/}{http://arohatgi.
info/}.}. In both cases the spectra were obtained using the Swedish–ESO 15~m Sub-millimetre Telescope (SEST) at La Silla, Chile with a beam size of 43$^{\prime\prime}$ at 115~GHz. The spectroscopic parameters of all the lines described above are summarised in Table~\ref{tab:spectroscopic_properties}. \\
We analysed the stability of the quoted continuum levels at 617\,GHz across scans (or time) and found the fluctuations to lie within 14\%. This is illustrated in Fig~\ref{fig:continuum_scans_scatter} and is not surprising as the observations were carried out using a wobbling secondary with a fast switching rate which guaranteed the removal of any drifts that may arise due to atmospheric instabilities. Furthermore, we find the 617\,GHz continuum flux to be well correlated with the continuum flux at 870\,$\mu$m observed using the Large APEX Bolometer Camera (LABOCA) at the APEX telescope presented in \cite{Wiess2008} which leads us to conclude that the continuum levels used are fairly reliable. 
\begin{table*}
    \centering 
    \caption{Spectroscopic properties of the studied species and transitions.}
    \begin{tabular}{lcclcrlr}
    \hline \hline
    Species & \multicolumn{2}{c}{Transition} & Frequency & $A_{\text{E}}$ & \multicolumn{1}{c}{$E_{\text{u}}$} & Receiver/Telescope & $\theta_{\rm FWHM}$\\
    &  $J^{\prime} - J^{\prime\prime}$ & $F^{\prime} - F^{\prime\prime}$ & [GHz] & [s$^{-1}$] & \multicolumn{1}{c}{[K]} & & \multicolumn{1}{c}{[$^{\prime\prime}$]}\\
    \hline 
    
    \arhp  & $1 - 0$  & --- & \phantom{0}617.5252(2) & 0.0045 & 29.63 & SEPIA660/APEX & 10.3\\
    \phtop & $3/2 - 1/2$& --- & \phantom{0}604.6841(8) & 0.0013 & 29.20 & SEPIA660/APEX & 10.3\\ 
     $N_{K_{\rm a}K_{\rm c}} = 1_{1,0} - 1_{0,1}$  & $3/2 - 3/2$&  --- & \phantom{0}607.2258(2) & 0.0062 & 29.20 & SEPIA660/APEX\\  
    \ohp & $2-1$ & $5/2 - 3/2$ & \phantom{0}971.8038(1)\tablefootmark{*} & 0.0182 & 46.64 & HIFI/Herschel & 22.0 \\ 
    $N = 1-0$& & $3/2 - 1/2$ & \phantom{0}971.8053(4) & 0.0152 & \\
         & & $3/2 - 3/2$ & \phantom{0}971.9192(11) & 0.0030 & \\ 
      & $1 - 1$ & $1/2 - 1/2$ & 1032.9985(7) & 0.0141  & 49.58 & 1.05THz Rx./APEX & 6.4\\
    &  & $3/2 - 1/2$ & 1033.0040(10) & 0.0035  &  \\
        & & $1/2 - 3/2$ & 1033.1129(7) & 0.0070  & \\
        & & $3/2 - 3/2$ & 1033.1186(10)\tablefootmark{*} & 0.0176  &  \\ 
     \ohtop & $3/2 - 1/2$ & $3/2 - 1/2$ & 1115.1560(8) &  0.0171 & 53.52 & HIFI/Herschel & 20.0\\
    $N_{K_{\rm a}K_{\rm c}} = 1_{1,1} - 0_{0,0}$ & & $1/2 - 1/2$ & 1115.1914(7) & 0.0274& \\
         & & $5/2 - 3/2$ & 1115.2093(7)\tablefootmark{*} & 0.0309& \\
         & & $3/2 - 3/2$ & 1115.2681(7) & 0.0138& \\
         & & $1/2 - 3/2$ & 1115.3035(8) & 0.0034& \\
    HF & $1-0$ & --- & 1232.4762(1) & 0.0242 & 59.14 & HIFI/Herschel & 17.0\\
    CO & $1-0$  & --- & \phantom{0}115.2712(0) & 7.203$\times10^{-8}$& 5.53 &  SEST & 43.0\\
    \hline 
    \end{tabular}
    \tablefoot{ The spectroscopic data are taken from the Cologne Database for Molecular Spectroscopy \citep[CDMS,][]{muller2005cologne}. The H$_2$O$^+$ frequencies were actually refined considering astronomical observations \citep[see Appendix A of][]{Mueller2016}, for which the upper level energies are given with respect to the ground state of p-H$_2$O$^+$ ($N_{K_{\rm a}K_{\rm c}} = 1_{0,1}$). For the rest frequencies, the numbers in parentheses give the uncertainty in the last listed digit. \tablefoottext{*}{Indicates the strongest hyperfine-structure transition, which was used to set the velocity scale in the analysis.}}
    \label{tab:spectroscopic_properties}
\end{table*}

\section{Results} \label{sec:results}
\subsection{Spectral line profiles} \label{subsec:spectral_profiles}
The calibrated and baseline-subtracted spectra towards NGC~253 and NGC~4945 are presented in Figure~\ref{fig:NGC253_allspectra_panel}. In the following paragraphs we qualitatively discuss the observed ArH$^+$ and p-H$_{2}$O$^+$ features and compare them to spectra of OH$^+$, o-H$_{2}$O$^+$, HF, CO, and H{\small I}. As mentioned above, our tuning setup simultaneously covers both the $J=3/2-1/2$ as well as the $J=3/2-3/2$ fine structure transitions from the ($1_{1,0}-1_{0,1}$) level of p-H$_{2}$O$^+$ at 604.684 and 607.225\,GHz. However, towards both NGC~253 and NGC~4945, we do not detect the $J=3/2-1/2$ fine structure transition near 604\,GHz above a noise level of ${\sim}$3 and 7\,mK at a spectral resolution of 4.5\,km~s$^{-1}$, respectively. Therefore this transition is not discussed any further in the text. 
For both NGC~253 and NGC~4945, we assume molecular source sizes of ${\sim}20^{\prime\prime}$, which were previously determined by \citet{Wang2004,Aladro2015,JP2018} and references therein, using molecular emission maps of abundant species like CO. The beam sizes of all the species studied here with the exception of CO (see Table~\ref{tab:spectroscopic_properties}) are either smaller than or comparable to this source size.

\subsubsection{NGC~253}
The \arhp line profile towards NGC~253 displays blueshifted absorption with respect to the systemic velocity of the source at 243\,km\,s$^{-1}$, covering a velocity range from ${\sim\!90}$ to 270\,km~s$^{-1}$ and centred at a velocity of 210\,($\pm 3.7$)\,km~s$^{-1}$. For comparison, other species like OH$^{+}$, \ohtop and HF all show spectra with P-Cygni profiles with the absorption seen at comparable blueshifted velocities. The corresponding H{\small I} profile shows an asymmetric absorption component, centred at a blueshifted velocity of ${\sim\!200}$\,km~s$^{-1}$. The velocity shift observed in the H{\small I} spectrum has been interpreted as evidence for a rotating nuclear disk of cold gas in this galaxy \citep{koribalski1995peculiar}. The P-Cygni profile observed towards the molecular ions studied here and HF, characterises the radial motion of the gas, which is indicative of outflows from the central region. The gas associated with the outflow component near 360~km~s$^{-1}$ is unsurprisingly not traced by ArH$^+$ and shows stronger emission in CO (particularly in its higher J-transitions) relative to the gas near 180~km~s$^{-1}$. Previous studies have determined the outflow component to be kinematically distinct from the surrounding gas and is associated with the so-called western-superbubble located north-west of the central starburst region \citep{Sakamoto2006} while the gas at 180~km~s$^{-1}$ is co-located with the central disk \citep{Krieger2019}. Such an asymmetric emission is revealed by H$\alpha$ observations whose emission is dominant on the approaching side of the outflow as reported by \citet{Bolato2013}.

\subsubsection{NGC~4945}
Towards NGC~4945, the \arhp line displays an asymmetric absorption profile between 455--715\,km~s$^{-1}$, similar to what is observed in the spectra of OH$^{+}$, \ohtop and HF lines. As discussed in \citet{monje2014hydrogen}, \citet{van2016ionization} and references therein, there are at least two velocity components, however unlike the other molecules and molecular ions, the broader component in \arhp is centred at $510\,$km~s$^{-1}$ and the narrower one at $605\,$km~s$^{-1}$ while for the other species the broader component is the redshifted absorption feature.  

In the p-H$_{2}$O$^+$, o-H$_{2}$O$^+$, OH$^{+}$\footnote{The \ohp spectrum contains a second absorption feature, which is the result of image band contamination from the CH$_{3}$OH ${J_k = 9_4-8_3~\text{E}2}$ line near 959.8\,GHz \citep{Xu1997}.}, and HF spectra, the two absorption components are observed to be almost symmetric about the galaxy's systemic velocity at 585\,km~s$^{-1}$ \citep{Chou2007}. This may indicate that the observed absorption arises from non-circular motions associated with the galaxy's bar. A similar absorption dip is seen in the emission line profiles of lines from higher density gas tracers like HCN, HCO$^+$, CN \citep{Henkel1990}, H$_{2}$CO \citep{Gardner1974} and also from CO \citep{Whiteoak1990} near 640\,km~s$^{-1}$. It likely traces foreground molecular gas that is moving towards the nucleus. 

The H{\small I} absorption spectrum against the nuclear continuum of the source, also displays a similar profile with two absorption components with dips at ${\sim}$540 and ${\sim}$620\,km~s$^{-1}$. The observed shift in the centres of both the \arhp absorption components in comparison to that of the other molecules, suggests that \arhp does not trace the same layers of infalling molecular gas as the other molecules and molecular ions but rather traces mostly or exclusively atomic gas layers, as expected. 

\subsubsection{PKS~1830$-$211}
As discussed in Sect.~\ref{sec:intro}, the $J=1-0$ transition of ArH$^+$ was first detected in extragalactic environments by \citet{Mueller2015} towards the intermediate redshift ($z=0.8858$) lensing galaxy located in front of the blazar, PKS~1830$-$211. These authors study absorption spectra extracted from two separate lines-of-sight towards two magnified images located on the south-west (SW) and north-east (NE) sides of its nucleus. Owing to differences in the nature of the continuum between PKS~1830$-$211's absorber and the sources studied here, a direct comparison of the spectral line profiles with those discussed in our study is not straightforward. While the background continuum for NGC~253 and NGC~4945 arise from the galaxies themselves, that for the sight lines studied towards foreground lensing galaxy in front of PKS~1830$-$211 arises from the distant quasar. Therefore, unlike the case for the nearby galaxies the spatial resolution for the latter is set by the small size of the background continuum emission (which is a few parsec in the plane of the $z$ = 0.8858 galaxy) but nonetheless we briefly describe the properties of the ArH$^+$ spectra observed towards both sight lines studied by \citet{Mueller2015}. The spectrum along the SW magnified image comprises of a single component with a line width of $\sim 57$~km~s$^{-1}$ and a weaker but broader blueshifted component, a feature that is also seen in the 607~GHz and 634~GHz p-H$_{2}$O$^+$ spectra. However, the spectral line profile of ArH$^+$ towards the NE image shows multiple narrow absorption features spanning 200~km~s$^{-1}$  which is comparable to the spread over which absorption from other diffuse gas tracers is seen in this direction, namely CH$^+$, HF, OH$^+$ and H$_2$O$^ +$ \citep{Mueller2016,Muller2017}.

\begin{figure*}
\centering 
    \includegraphics[width=0.404\textwidth]{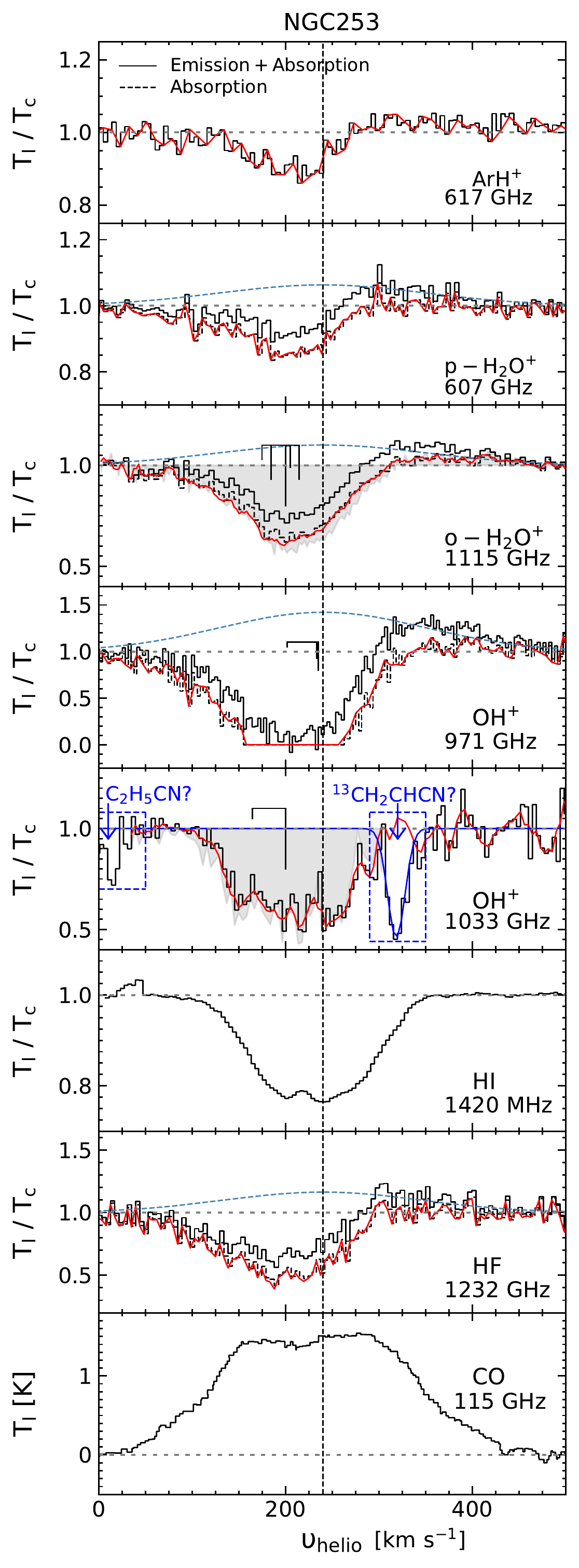}\quad 
    \includegraphics[width=0.404\textwidth]{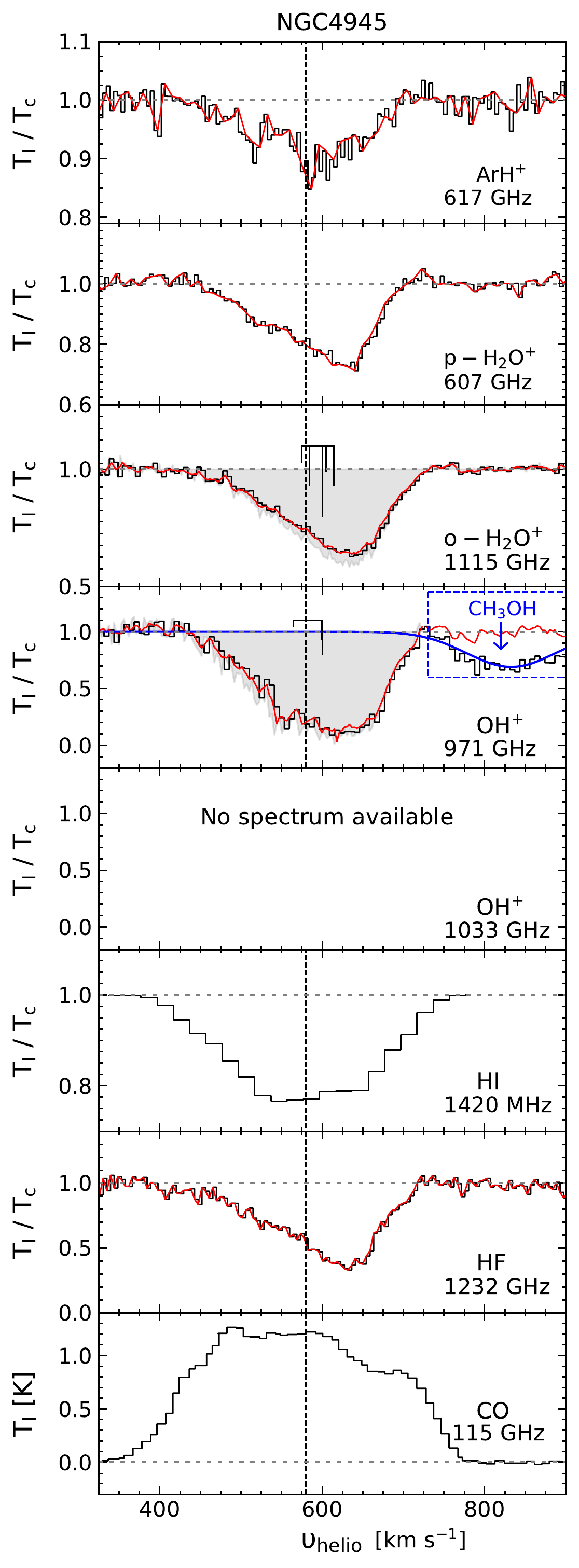}
    \caption{From top to bottom: Normalised absorption spectra of ArH$^+$, p-H$_{2}$O$^+$, o-H$_{2}$O$^+$, OH$^+$ (at 971 and 1033~GHz), H{\small I} and HF as well as the CO (1-0) emission line spectrum for comparison, towards NGC~253 (left) and NGC~4945 (right), respectively. In the spectra for NGC~253, the dotted light blue curves display the Gaussian fit to the emission component. The dashed black line represents the individual absorption profiles after subtracting the Gaussian fit with their Wiener filter fits overlaid in red for all species except H{\small I} and CO. The relative intensities of the hyperfine structure (HFS) components of the o-H$_{2}$O$^+$ and OH$^+$ transitions are shown in black above their respective spectra and the grey shaded regions display their HFS deconvolved spectra. The vertical dashed black lines mark the systemic velocity of NGC~253 and NGC~4945 at $240\,$km~s$^{-1}$ and $563\,$km~s$^{-1}$, respectively. The 971~GHz OH$^+$ line profile towards NGC~253 is saturated at blueshifted velocities while the 1033~GHz OH$^+$ spectrum is potentially contaminated by the $^{13}$CH$_{2}$CHCN (21$_{10,11}$--20$_{9,12}$) line at 1032.783~GHz (fit shown by the dark blue curve). Also marked here in blue is contamination from the C$_{2}$H$_{5}$CN ($37_{9,29}$--$37_{6,32}$) line at 1033.868~GHz. The 971~GHz OH$^+$ spectrum towards NGC~4945 is contaminated by the  CH$_{3}$OH ${J_k = 9_{4,6}-8_{,5}3~\text{E}}$ line near 959.900~GHz originating from the image sideband (marked in blue). }
    \label{fig:NGC253_allspectra_panel}
\end{figure*}

\subsection{Column densities}\label{subsec:column_densities}
The line profiles can be expressed in terms of the optical depth, $\tau$, using the radiative transfer equation, which for the particular case of absorption spectroscopy is given by $T_\text{b} = T_\text{c}e^{-\tau}$, where $T_{\text{b}}$ and $T_{\text{c}}$ are the line, and the background continuum brightness temperatures, respectively. For NGC~253, we compute optical depths for the absorption profile obtained after subtracting the emission component. The emission component is modelled by fitting a Gaussian profile centred at the systemic velocity of the source. We fitted the optical depth profile on heliocentric velocity scales, that is to say $\tau$ versus. $\upsilon_{\rm helio}$, using the Wiener filter fitting technique as described in \citet{jacob2019fingerprinting}. This fitting procedure first fits the spectrum using the Wiener filter kernel by minimising the mean square error between the model ($T_\text{c}e^{-\tau}$) and observations. For species like OH$^+$ and o-H$_{2}$O$^{+}$, whose rotational transitions further undergo hyperfine structure (HFS) splitting, this algorithm additionally deconvolves the HFS from the observed spectrum using the relative spectroscopic weights of the different HFS components. When fitting lines that do not exhibit HFS, the procedure simply assumes that there is only a single HFS component whose frequency corresponds to that of the fine-structure transition itself. Other than the observed spectrum and the spectroscopic parameters of the line to be fit, the only other input parameter required by the Wiener filter technique is the spectral noise, which is assumed to be independent of the observed signal. However, the Wiener filter faces singularities in portions of the spectrum in which the observed line profiles saturate or the line-to-continuum ratio tends to zero. This is the case for the OH$^+$ spectrum towards NGC~253, which shows saturated absorption at blueshifted velocities between 185 and 235\,km~s$^{-1}$ (see Fig.~\ref{fig:NGC253_allspectra_panel}). Therefore, as discussed in Sect.~\ref{sec:observations} we instead model the 1033~GHz transition of OH$^+$. Prior to performing HFS deconvolution we remove contributions from the $^{13}$CH$_{2}$CHCN contamination using the WEEDS package in the GILDAS software.

\begin{figure*}
    \centering
    \includegraphics[width=0.44\textwidth]{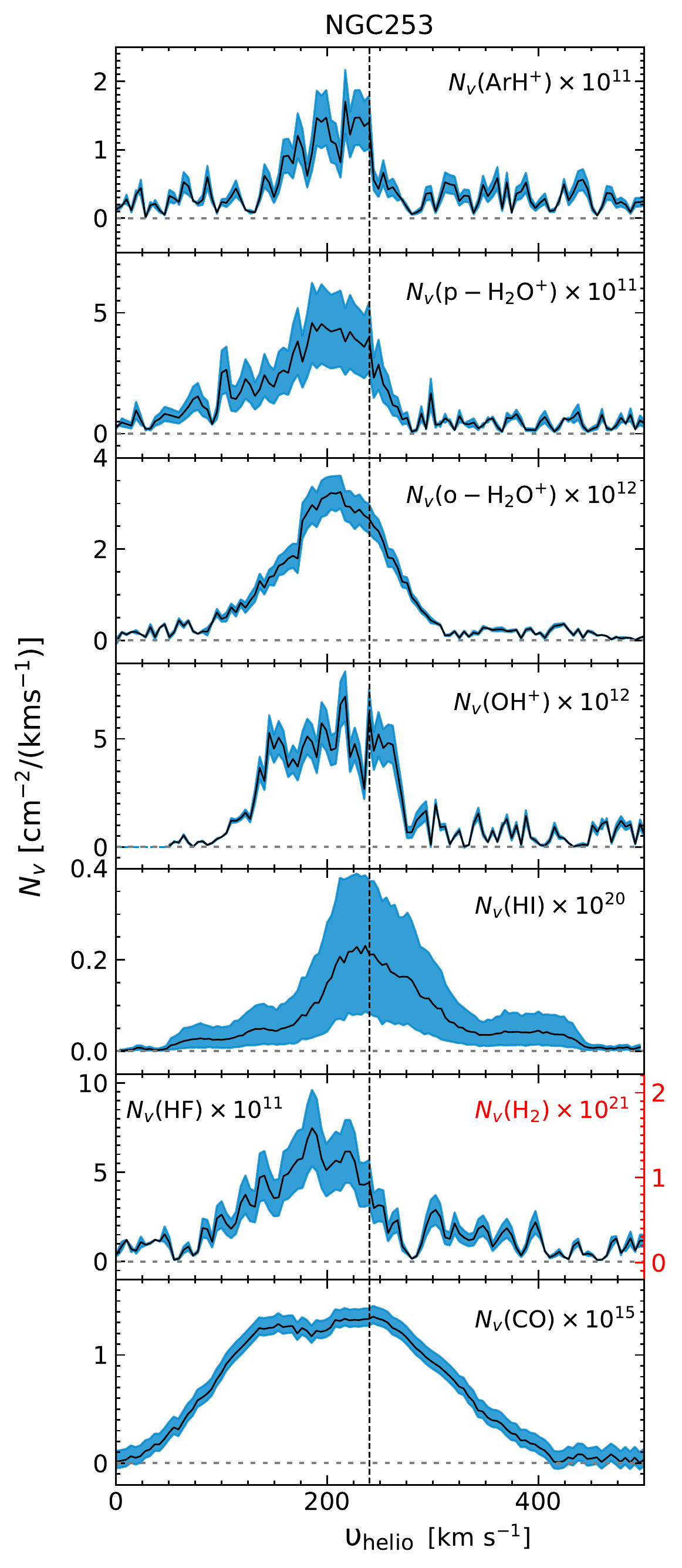}\quad\includegraphics[width=0.455\textwidth]{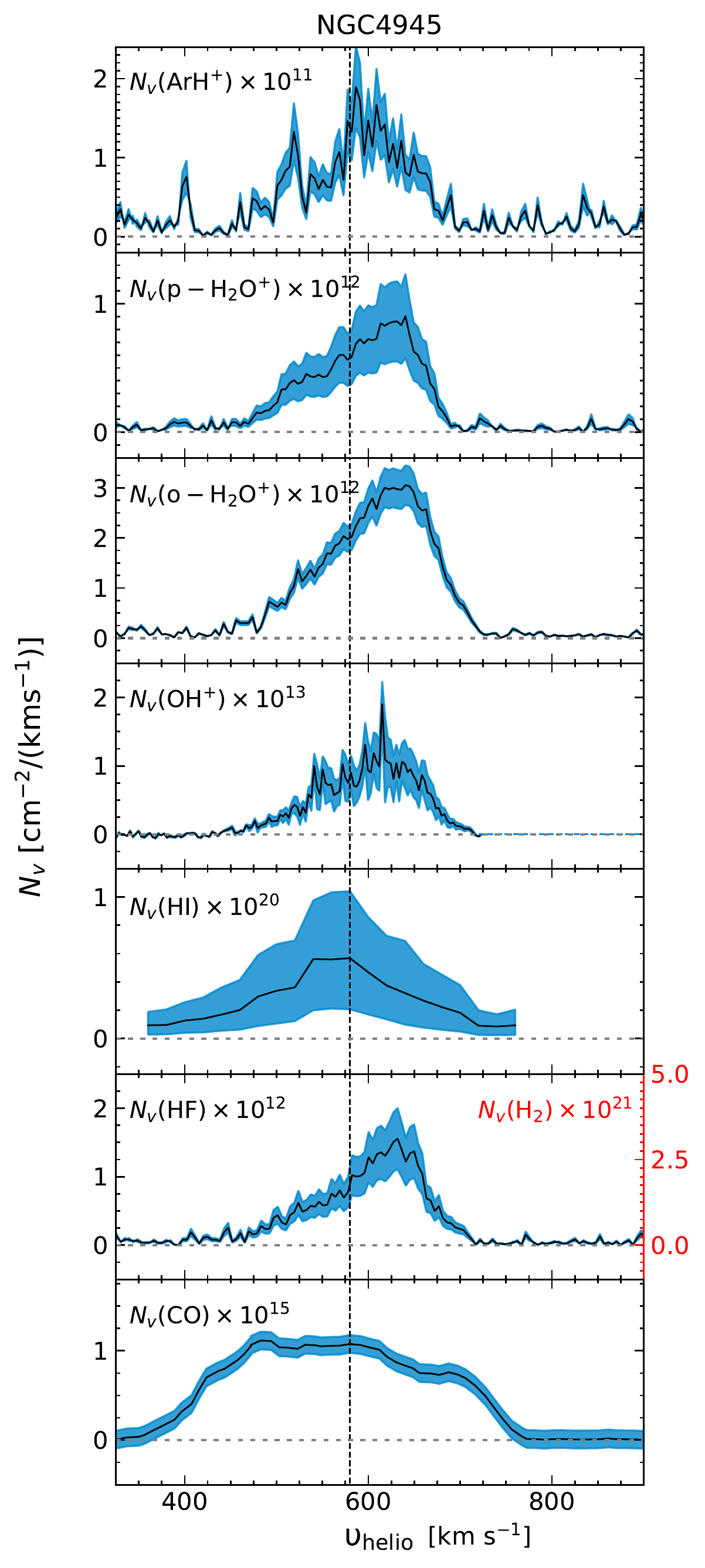}
    \caption{From top to bottom: Column density distributions (black) of ArH$^+$, p-H$_{2}$O$^+$, o-H$_{2}$O$^+$, OH$^+$, H{\small I}, HF alongside the corresponding scaled H$_{2}$ profile ([HF]/[H$_{2}$] = 5$\times 10^{-10}$; \citet{Emprechtinger2012}) and CO towards NGC~253 (left) and NGC~4945 (right), respectively. The corresponding uncertainties are displayed by the blue shaded region.}
    \label{fig:coldens_distributions}
\end{figure*}

The resulting optical depth profiles ($\tau_{\rm decon}$ versus. $\upsilon_{\rm helio}$) are used to derive column densities as follows, assuming that the foreground absorption covers the background continuum source entirely:
\begin{equation}
    N_{\text{tot}} = \frac{8\pi\nu^3 }{b_{\text{ff}}c^3 } \frac{Q(T_{\text{ex}})}{g_{\text{u}} A_{\text{E}}} \text{e}^{E_{\text{u}}/T_{\text{ex}}} \left[ \text{exp} \left(\frac{h\nu}{k_{\text{B}}T_{\text{ex}}}\right) - 1 \right]^{-1} \int \tau_{\rm decon}\text{d}\upsilon \, .
    \label{eqn:column_density}
\end{equation}
For a given species, all the spectroscopic terms in Eq.~\ref{eqn:column_density} namely, the upper level energy, $E_{\text{u}}$, the upper level degeneracy, $g_{\text{u}}$, and the Einstein A coefficient, $A_{\text{E}}$, all remain constant, except for the partition function, $Q$, which itself is a function of the rotation temperature, $T_{\text{rot}}$. Under conditions of local thermodynamic equilibrium (LTE), $T_{\text{rot}}$ is equal to the excitation temperature, $T_{\text{ex}}$. The excitation of the molecules is straightforward as most of the particles  are expected to occupy the ground state level, owing to the large Einstein A coefficients of all the lines involved, resulting in high critical densities of the order of a few 10$^{7}$\,cm$^{-3}$ (see Table~\ref{tab:spectroscopic_properties} for the Einstein A coefficients). Therefore, assuming a complete ground state occupation, we can approximate the excitation temperature to be less than the energy of the upper level above the ground and equal to the radiation temperature of the cosmic microwave background ($T_{\rm CMB} = 2.73\,$K). Similar assumptions for the excitation temperatures have been made by \citet{van2016ionization} in their analysis. Since, ${T_{\rm CMB} < T_{\rm ex} < E_{\rm u}/k_{\rm B}}$, the column densities derived for the different species studied here strictly represent only lower limits. 
Unlike those of other species, the CO column densities are determined from the integrated intensities of their respective emission line profiles, for an excitation temperature of 20~K, as was assumed by \citet{Houghton1997} and \citet{Curran2001}. Furthermore, the derived column densities are corrected to a first order for beam dilution effects, using the beam filling factor, $b_{\text{ff}}$, following $b_{\text{ff}} = \left[ \left( \theta_{\text{s}}^2 + \theta_{\text{b}}^2 \right)/\theta_{\text{s}}^2 \right]$ where, $\theta_{\text{s}}$, and $\theta_{\text{b}}$ represent the molecular source size and the beam size, respectively. The column density profiles hence derived per velocity channel are displayed in Fig.~\ref{fig:coldens_distributions} and the total column densities derived by integrating between 55 and 295~km~s$^{-1}$ for NGC~253, and 445 and 725~km~s$^{-1}$ for NGC~4945, are summarised in Table~\ref{tab:col_dens}. Using the column densities we determine from our spectra of the 607\,GHz p-H$_{2}$O$^+$ line, we predict, assuming conditions of LTE at 2.73\,K, integrated intensities of 1.3 and 2.9\,K\,km\,s$^{-1}$ for the 604\,GHz transition of p-H$_{2}$O$^+$, values that are lower than our 3$\sigma$ upper limits of 2.1 and 5.2\,K\,km\,s$^{-1}$ towards NGC~253 and NGC~4945, respectively.

\begin{table*}
\caption{Synopsis of the derived column densities. }
\centering 
    \begin{tabular}{lr rrrrrr}
    \hline \hline
        \multicolumn{1}{c}{Source} & $\upsilon_{\text{min}}$--$\upsilon_{\text{max}}$ & \multicolumn{1}{c}{$N(\text{ArH}^{+})$} &  \multicolumn{1}{c}{$N(\text{p-H}_{2}\text{O}^{+})$} & \multicolumn{1}{c}{$N(\text{o-H}_{2}\text{O}^{+})$} &  \multicolumn{1}{c}{$N(\text{OH}^{+})$}  & \multicolumn{1}{c}{$N(\text{HF})$}  & \multicolumn{1}{c}{$N(\text{H{\small I}})$} \\ 
        & $~$[km~s$^{-1}$] & \multicolumn{1}{c}{10$^{12}$[cm$^{-2}$]} & \multicolumn{1}{c}{10$^{13}$[cm$^{-2}$]} & \multicolumn{1}{c}{10$^{13}$[cm$^{-2}$]} & 
        \multicolumn{1}{c}{10$^{14}$[cm$^{-2}$]} & \multicolumn{1}{c}{10$^{14}$[cm$^{-2}$]} & \multicolumn{1}{c}{10$^{20}$[cm$^{-2}$]}\\
         \hline 
         NGC~253  & 55--295  & $3.35 \pm 0.31$ & $1.22 \pm 0.22$  &  $ 8.47 \pm 0.36$  & \multicolumn{1}{c}{>1.57} & $1.40 \pm 0.60$ & $4.57 \pm 3.80 $\\
         NGC~4945 & 445--725 & $4.31 \pm 0.20$ & $2.46 \pm 0.24$ & $9.28 \pm 0.32$ & \multicolumn{1}{c}{>3.20} & $1.61 \pm 0.68$ & $4.27 \pm 3.93$\\
         \hline 
    \end{tabular}
    
    \label{tab:col_dens}
\end{table*}

\subsection{Cosmic-ray ionisation rates}\label{subsec:CRIR}
Cosmic-rays represent the dominant source of heating and ionisation in the inner parts of molecular clouds that are not penetrated by UV photons, making them an important driver for ion-molecular reactions, including those responsible for the formation of ArH$^+$ \citep{roach1970potential}:

\begin{equation*}
  \text{Ar}  \xrightarrow{\text{CR}} \text{Ar}^{+} + e^{-}\xrightarrow{\text{H}_{2}}
     \text{ArH}^{+} + \text{H} \hspace{2.25cm} (\Delta E = 1.436~\text{eV})\,. \\
\end{equation*}
Hence, as a key ingredient in the ensuing chemistry, the abundance of \arhp is sensitive to the cosmic-ray ionisation rate, $\zeta_{\text p}(\text{H})$, and to the molecular fraction, $f_{\text{H}_{2}}$, of the gas probed. The cosmic-ray ionisation rates in  both NGC~253 and NGC~4945 have previously been determined by analysing the steady-state chemistry of oxygen-bearing ions like OH$^+$ and H$_{2}$O$^+$ by \citet{van2016ionization}. 
Following the steady-state analysis presented by these authors and \citet{indriolo2015herschel}, we derive revised cosmic-ray ionisation rates by including contributions from p-H$_{2}$O$^+$. The value of $N_{\text H}$ used in our calculations is given by $N$(H{\small I}) + 2$\times N$(H$_{2}$), where $N$(H$_{2}$) is obtained by using $N$(HF) as a surrogate for molecular hydrogen, where the abundance of HF in dense gas 
is assumed to be 5$\times10^{-10}$ \citep{Emprechtinger2012}. From the column density profiles presented in Fig.~\ref{fig:coldens_distributions}, it is clear that the total hydrogen content along the studied sight lines are dominated by dense molecular gas with pockets or bubbles of diffuse clouds as suggested by \citet{van2016ionization}. Including contributions from p-H$_{2}$O$^+$ we derive mean cosmic-ray ionisation rates of $2.2\times10^{-16}$ and $7.5\times10^{-17}$\,s$^{-1}$ across NGC~253 and NGC~4945, respectively, values that are not far from those derived by \citet{van2016ionization}. The derived cosmic-ray ionisation rates represent only lower limits because of uncertainties in the derived column densities, particularly in that of the H{\small I} and OH$^+$ absorption profiles. In addition other assumptions made in this calculation, such as the values of the gas density ($n_{\rm H} = 35\,$cm$^{-3}$), the electron fraction ($x_{\rm e}=1.5\times10^{-4}$) and the OH$^+$ formation efficiency parameter ($\epsilon=7\%$) are poorly constrained in these sources. The general impact of the uncertainties associated with the assumptions made in such an analysis, on the derived cosmic-ray ionisation rates of Galactic sources is discussed in more detail in \citet{schilke2010} and \citet{indriolo2015herschel}.

\subsection{Gas properties}\label{subsec:gas_properties}

The cosmic-ray ionisation rates discussed in the previous section, $\zeta_{\rm p}({\rm H})$, describe only the total number of primary ionisations per hydrogen atom, per second. In contrast, the total cosmic-ray ionisation rate, $\zeta_{\rm t}$, includes contributions from both primary ionisation by cosmic-rays as well as ionisation by secondary electrons (resulting from the former). As discussed by \citet{Neufeld2017}, the exact relation between the two rates is dependent on the molecular fraction and fractional ionisation, but for the typical conditions of diffuse clouds the two are roughly related as $\zeta_{\rm p}(\rm H)=\zeta_{\rm t}/1.5$. Therefore not only do cosmic-rays play an important role in heating the gas, they also increase the electron abundance.

In the following sections we evaluate the impact of cosmic-ray ionisation and heating through the physical properties traced by ArH$^+$. We estimate the molecular fraction characteristic of the gas traced by ArH$^+$ and explore the impact of collisional excitation by electrons on the observed ArH$^+$ abundance.

\subsubsection{Molecular fraction}\label{subsubsec:molec_frac}
Under the assumption that the cloud volumes containing ArH$^+$ are exposed to the same cosmic-ray ionisation flux as those traced by both OH$^+$ and H$_{2}$O$^+$, we can derive the molecular fraction, $f_{\text{H}_{2}}$, of the gas probed by ArH$^{+}$, using the \arhp abundances (with respect to H{\small I}) derived from observations as a constraint. We first present the relation between $\zeta_{\text{p}}$(H), $X({\text{ArH}}^+)$ and $f_{\text{H}_{2}}$ by analysing the steady-state ion-molecular chemistry of \arhp as discussed in \citet{schilke2014ubiquitous}. \arhp is destroyed primarily via proton transfer reactions with H$_{2}$ or atomic oxygen, and photodissociation:  
\begin{align*}
    \text{ArH}^{+} &+ \text{H}_{2} \rightarrow \text{Ar} + \text{H}_{3}^{+} \, ; \quad \quad \quad k_{5} = 8\times 10^{-10}~\text{cm}^{3}\text{s}^{-1} \\
    \text{ArH}^{+} &+ \text{O} 
    \rightarrow \text{Ar} + \text{OH}^{+} \, ; \quad  \quad \,\,\, \, k_{6} = 8\times 10^{-10}~\text{cm}^{3}\text{s}^{-1} \\ 
    \text{ArH}^{+} &+ \text{h}\nu \rightarrow \text{Ar}^+ + \text{H} \, ; \, \quad \quad \quad k_{7} = 1.0\times 10^{-11}\chi_\text{UV}f_{\text{A}}~\text{s}^{-1} \, .
\end{align*}
The photodissociation rate of \arhp was estimated by \citet{alekseyev2007theoretical} to be $\sim\!1.0\times10^{-11}f_{\text{A}}\,\text{s}^{-1}$ for an unshielded cloud model that is uniformly surrounded by the standard Draine UV interstellar radiation field. 
The attenuation factor, $f_{\text{A}}$, is given by an exponential integral and is a function of visual extinction, $A_\text{v}$. For a cloud model with $A_{\text{v}} = 0.3$, \citet{schilke2014ubiquitous} derived values for $f_{\text{A}}$ between 0.30 and 0.56 that increase as you move outwards from the centre of the cloud. For our analysis, we use 0.43 for $f_{\text{A}}$, which lies mid-way through the computed range of values. 

\begin{figure*}
\includegraphics[width=0.48\textwidth]{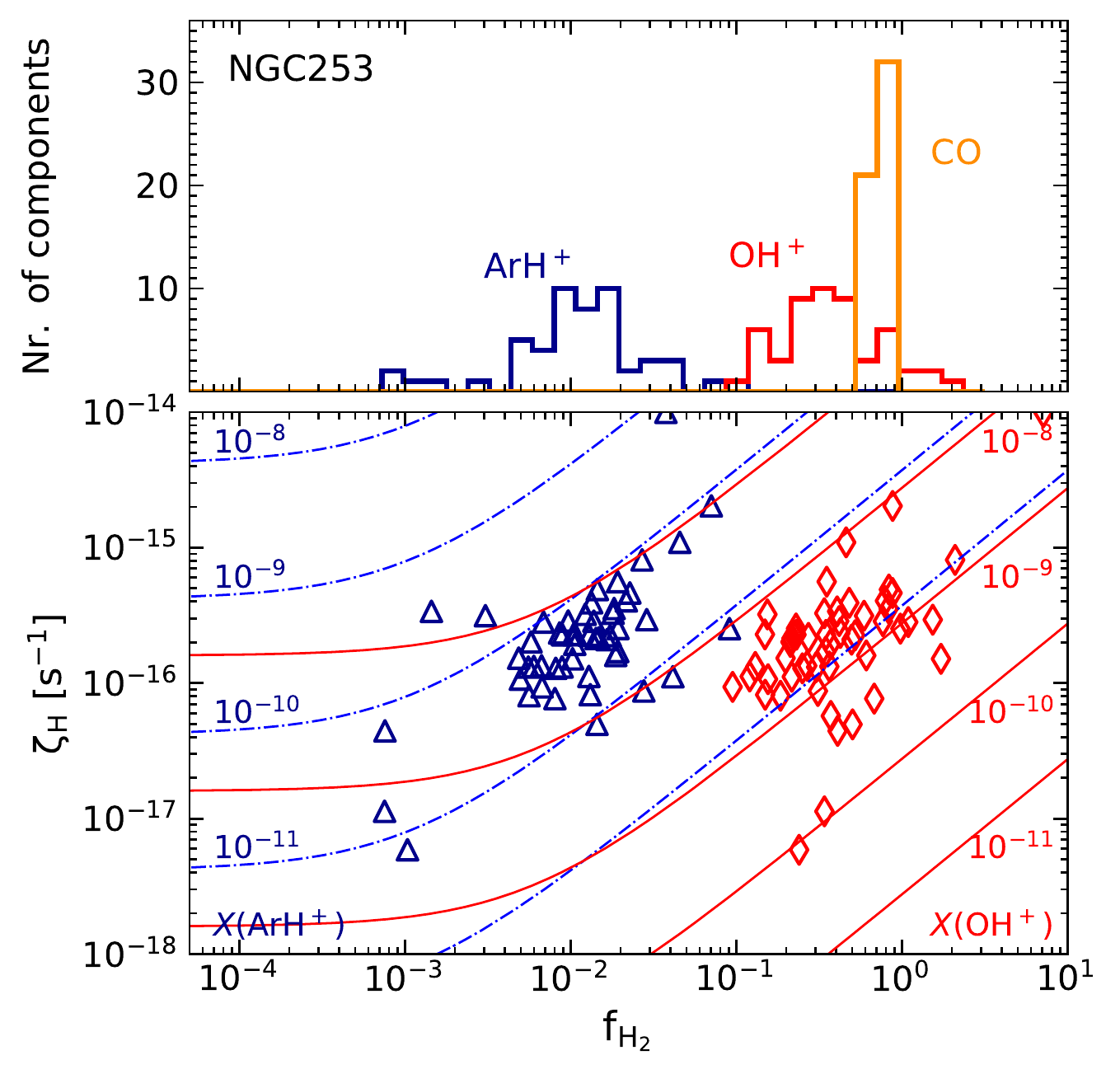}\quad
\includegraphics[width=0.48\textwidth]{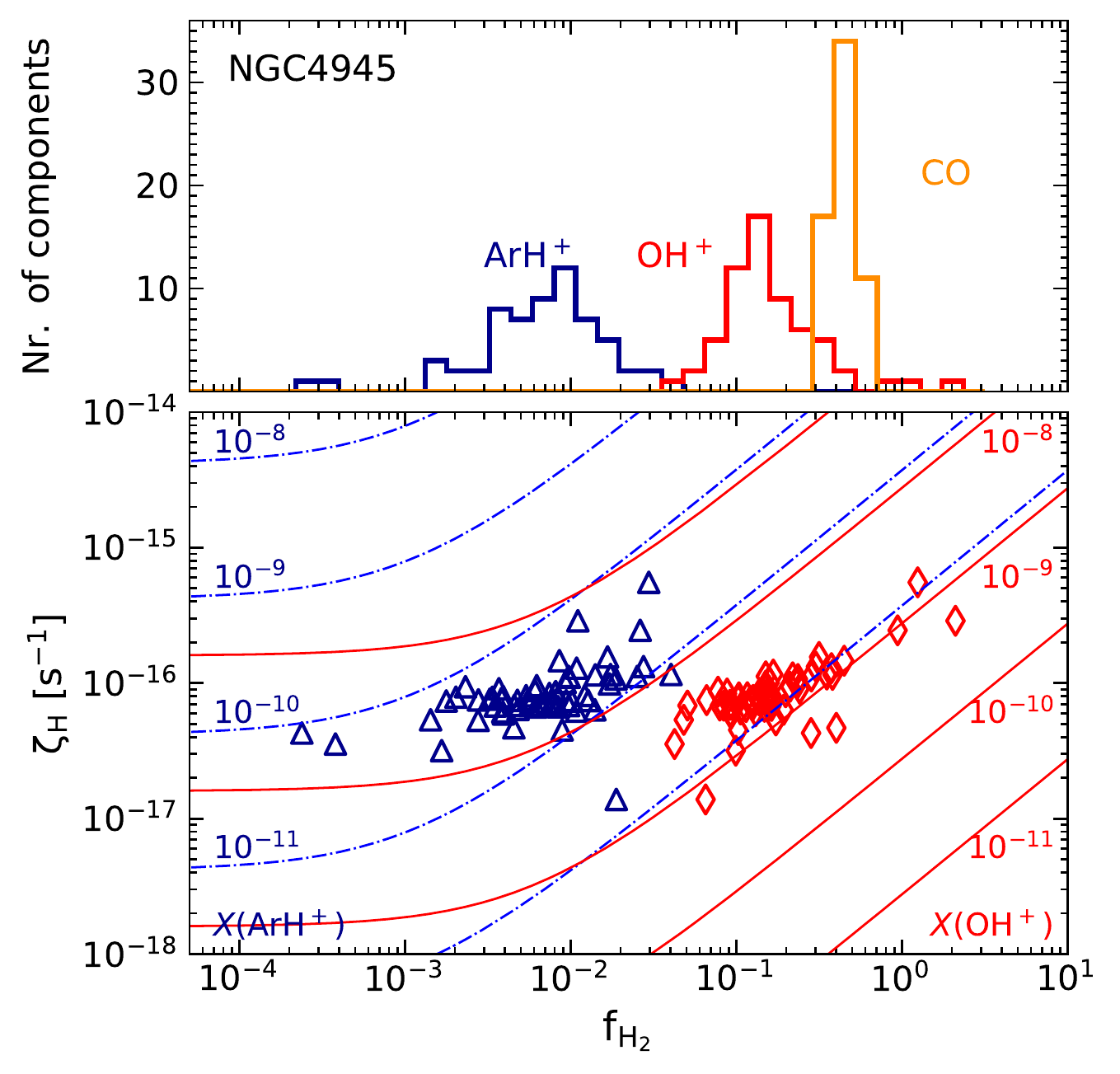}
\caption{Top: Distribution of molecular gas fraction ($f_{\text{H}_{2}}$) traced by ArH$^+$ (blue), OH$^+$ (red) and CO (orange). Bottom: Contours of $X(\text{ArH}^+)$ (blue) and $X(\text{OH}^+)$ (red) abundances with respect to $N_{\text{H}}$ in the $f_{\text{H}_{2}}$--$\zeta_{\rm p}{\text{(H)}}$ plane. Blue triangles and red diamonds represent the corresponding values derived from the LOS observations presented in this work computed per channel in the velocity interval (${\Delta\upsilon \sim\!4.5}\,$km~s$^{-1}$) between 55 and 295\,km\,s$^{-1}$, and 445 and 725\,km\,s$^{-1}$ for NGC~253 and NGC~4945, respectively.}
\label{fig:gas_prop}
\end{figure*}

We further assume an atomic oxygen abundance (relative to H nuclei) of ${3.9 \times 10^{-4}}$ \citep{Cartledge2004} and an argon abundance close to this element's solar abundance of ${3.2\times 10^{-6}}$ \citep{lodders2008solar}. Observations by \citet{van2016ionization} suggest that the dense and diffuse gas phases are well mixed in the galaxies under consideration here, a notion based on the similarities between the observed profiles of OH$^+$ and H$_{2}$O$^+$ lines to those of H$_{2}$O and H{\small I} lines. Therefore, akin to these authors, who derived the cosmic-ray ionisation rates for NGC~253 and NGC~4945 by analysing the steady-state chemistry of OH$^+$ and H$_{2}$O$^+$, we assume a gas density, $n(\text{H})$, of $35\,\text{cm}^{-3}$ \citep{indriolo2015herschel} for the OH$^+$--H$_{2}$O$^+$ absorbing clouds in our analysis.   
\noindent 
Using these values, the cosmic-ray ionisation rate can be approximated as
\begin{align}
    \zeta_\text{p}(\text{H}) ~({\rm s}^{-1}) &= \frac{N(\text{ArH}^{+})}{N_{\text{H}}}\left(\frac{k_{5}n(\text{H}_{2}) + k_{6}n(\text{O}) + k_{7}}{11.42}\right)\,({\rm s}^{-1}) ,\\ &=\frac{N(\text{ArH}^{+})}{N_{\text{H}}}\left( \frac{0.5005 + 448f_{\text{H}_{2}}}{1.2 \times 10^{6} }\right) \, .
    \label{eqn:CRIR_arhp}
\end{align}
Eq.~\ref{eqn:CRIR_arhp} is re-arranged and expressed in terms of $f_{\text{H}_{2}}$, as
\begin{equation}
  f_{\text{H}_{2}}  = 2.68\times10^{3}\,({\rm s}) \,\left[ \frac{\zeta_\text{p}(\text{H})}{X(\text{ArH}^{+})} - 4.17\times10^{-7} \right]\,({\rm s}^{-1}) \, . 
  \label{eqn:hfrac_arhp}
\end{equation}
Using Eq.~\ref{eqn:hfrac_arhp}, the molecular fractions of the gas probed by ArH$^{+}$ are found to be a few times ${10^{-3}}$ towards both sources, values that are comparable to what is derived along the NE line-of-sight towards PKS~1830$-$211 (which amongst the two sight lines studied by \citet{Mueller2015} has been shown to probe more diffuse and atomic environments \citep{Koopmans2005}) with $X$(ArH$^+$) = $2.8\times10^{-9}$ and $\zeta_\text{p}(\text{H}) = 3\times10^{-15}$~s$^{-1}$. The molecular gas fraction hence derived, per velocity channel over velocity intervals most relevant to the sources studied (as discussed in Table~\ref{tab:col_dens}) is displayed in Fig.~\ref{fig:gas_prop}. Towards both sources we find the ArH$^+$ bearing clouds to host abundances between 10$^{-11}$ and 10$^{-10}$ and trace molecular gas fractions between 3$\times10^{-4}$ and a few times {$10^{-2}$}. In comparison, OH$^+$, whose abundance is almost two orders of magnitude higher than that of ArH$^+$, mainly traces gas with molecular fractions of the order of a few times 10$^{-1}$ whereas \citet{Mueller2016} derived $f_{{\rm H}_{2}}$ values for OH$^+$-bearing gas that varies between 0.01 and 0.07 towards both sight lines in their study. Therefore, the different atomic gas tracers may not be spatially co-existent. Furthermore, the transition from atomic to molecular gas is illustrated in the top panel of Fig.~\ref{fig:gas_prop} through the distribution of the molecular gas fractions traced by ArH$^+$-, OH$^+$-, and CO-bearing gas volumes. The molecular fractions in this analysis are computed using Eq.~\ref{eqn:hfrac_arhp}, Eq.~12 of \citet{indriolo2015herschel} and by assuming a CO-to-H$_{2}$ conversion factor of 2$\times10^{20}\,$cm$^{-2}$\,(K\,km\,s$^{-1}$)$^{-1}$ as recommended by \citet{Bolatto2013}, albeit with a factor of 2 uncertainty. By combining previous estimates of the cosmic-ray ionisation rates derived for Arp~220, of $\zeta_{\rm p}$(H)>$10^{-13}~$s$^{-1}$ \citep{Gonzalez2013,van2016ionization}, with our upper limit for the \arhp abundance of 2.5$\times10^{-11}$ (determined by integrating the $3\sigma$ detection limit quoted in Sect.~\ref{sec:observations} over a line width of 743~km~s$^{-1}$ \citep{Mirabel1982} and $N$(H$_{2}) = 1\times10^{24}~$cm$^{-2}$ \citep{Downes2007}), we are unable to derive reasonable values for the molecular fraction traced by ArH$^+$.  

\subsubsection{Excitation by electrons}\label{subsubsec:electron_frac}
In this section we explore the effects of collisional excitation of ArH$^+$ by electrons. We expect ArH$^+$ to reside in cloud layers with electron fractions, $x_{\rm e}$ of $\geq 10^{-5}$--$10^{-4}$, making them competitive collision partners to atomic hydrogen in such cloud environments. Moreover, the destruction of ArH$^+$ via dissociative recombination reactions with electrons have, alongside photodisocciation pathways, been shown to have small to negligible impacts on the ArH$^+$ abundances in astrophysical environments \citep{alekseyev2007theoretical, Mitchell2005, Abdoulanziz2018}, making electron-impact excitation effects of significant importance for ArH$^+$. In order to explore this and to evaluate the validity of our assumptions for the electron density and excitation temperature, we performed non-LTE radiative transfer calculations using the statistical equilibrium radiative transfer code RADEX \citep{vanderTak2007}. The models were run for a uniform sphere geometry (with the offline version of RADEX, which allows a choice of input geometries) 
under the large velocity gradient (LVG) approximation. Based on the escape probability formalism, the code computes level populations, line intensities, excitation temperatures, and optical depths as a function of the physical conditions (kinetic temperature and density) and radiative transfer parameters (column density and line width) specified as inputs. The models were run using rate coefficients recently computed by \citet{Hamilton2016} for collisions between \arhp and electrons as well as those between \arhp and atomic hydrogen computed by \citet{Dagdigian2018}. 

\begin{figure}
    \centering
    \includegraphics[width=0.5\textwidth]{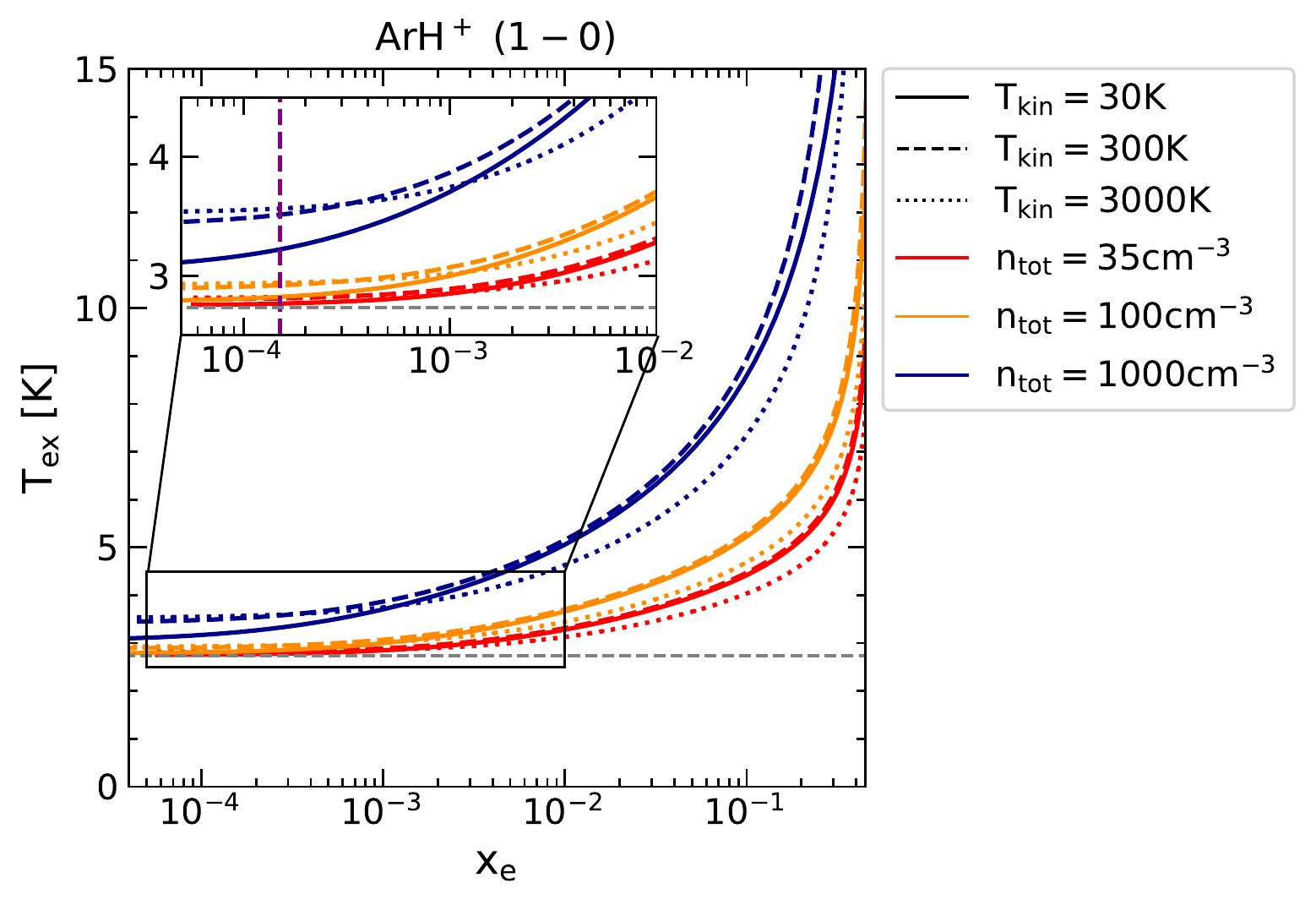}
    \caption{RADEX modelled excitation temperature as a function of electron density, for fixed values of the total gas density at 35 (red), 100 (orange) and 1000\,cm$^{-3}$ (dark blue) for the ArH$^+$ (1-0) transition. Furthermore, each model is run for fixed gas temperatures at 30 (solid), 300 (dashed) and 3000\,K (dotted), respectively. The horizontal dashed grey line marks an excitation temperature of 2.73\,K. The inset zooms in on 10$^{-4} < x_{\rm e}<10^{-2}$ with the vertical dashed purple line indicating $x_{\rm e} = 1.5\times10^{-4}$ which corresponds to the value of $x_{\rm e}$ used in our calculations. }
    \label{fig:arhp_tex}
\end{figure}

By constraining the models using the \arhp column densities as determined in Sect.~\ref{subsec:column_densities}, we model the excitation temperature of the observed 617\,GHz ArH$^+$ line as a function of the electron density and gas temperature, $T_{\rm kin}$. The models are run across a 100$\times$100 grid, with $x_{\rm e}$ values between 10$^{-6}$, and 10$^{-1}$ and gas temperatures between 10 and 3000\,K. Furthermore, the total gas density, $n_{\rm H}$, approximated by the sum of the electron and atomic gas densities in the models, is fixed to values of 35, 100 and 1000\,cm$^{-3}$. 
The modelled results are visualised in Fig.~\ref{fig:arhp_tex}. Similar to \citet{Dagdigian2018}, we find the excitation temperature for the models with total gas densities fixed at 100 and 1000\,cm$^{-3}$ to be greater than our assumed value of $T_{\rm ex}$ at 2.73\,K, across the entire $x_{\rm e}$ parameter space, with values of $T_{\rm ex}$ only slightly higher than 2.73 for $x_{\rm e} \lessapprox 10^{-3}$, beyond which the value of the excitation temperature increases quite rapidly. While the general trend for those models with a total gas density of 35\,cm$^{-3}$ (equal to the gas densities assumed in Sect.~\ref{subsec:CRIR} to compute the cosmic-ray ionisation rates) is similar to those with higher gas densities, that is, 100 and 1000\,cm$^{-3}$, the ArH$^+$ excitation temperature over the assumed electron fraction ($\approx1.5\times10^{-4}$ see Sect.~\ref{subsec:CRIR}) remains very close to a value of 2.73\,K. This implies electron densities of $\sim\!5.3\times 10^{-3}$, consistent with the low electron abundances expected in gas volumes bearing OH$^+$, a species formed in cloud environments similar to those that contain  ArH$^+$. Low electron densities are prerequisites for the formation of detectable amounts of OH$^+$, whose formation pathway via H$_{3}^+$ and atomic oxygen competes with the dissociative recombination of H$_{3}^+$ with electrons. Observationally however, the range of electron densities is poorly constrained with a possible upper limit set by assuming the electron densities to be consistent with the photoionisation of neutral carbon. 

\begin{figure*}
    \centering
     \includegraphics[width=0.48\textwidth]{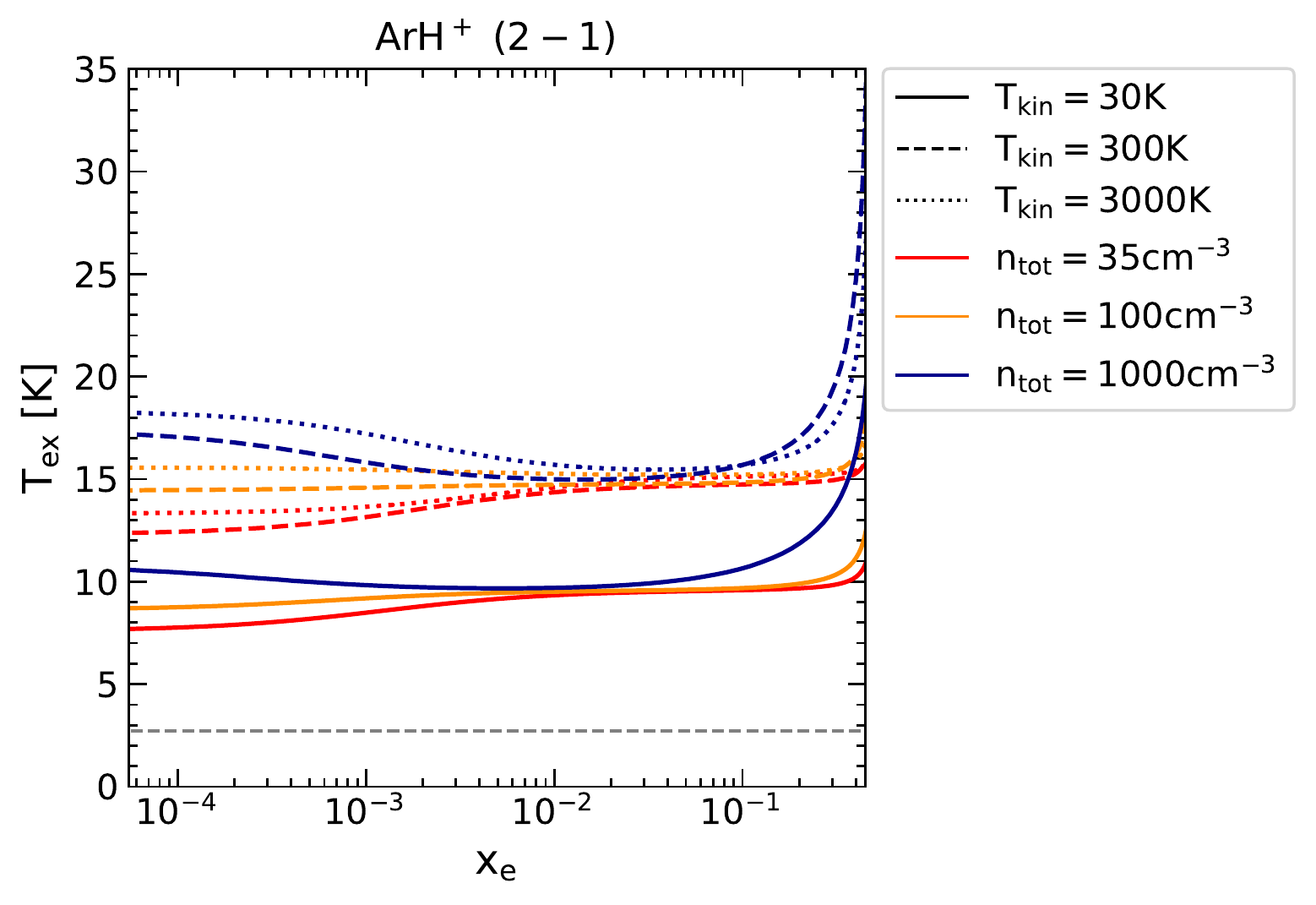} \quad
    \includegraphics[width=0.48\textwidth]{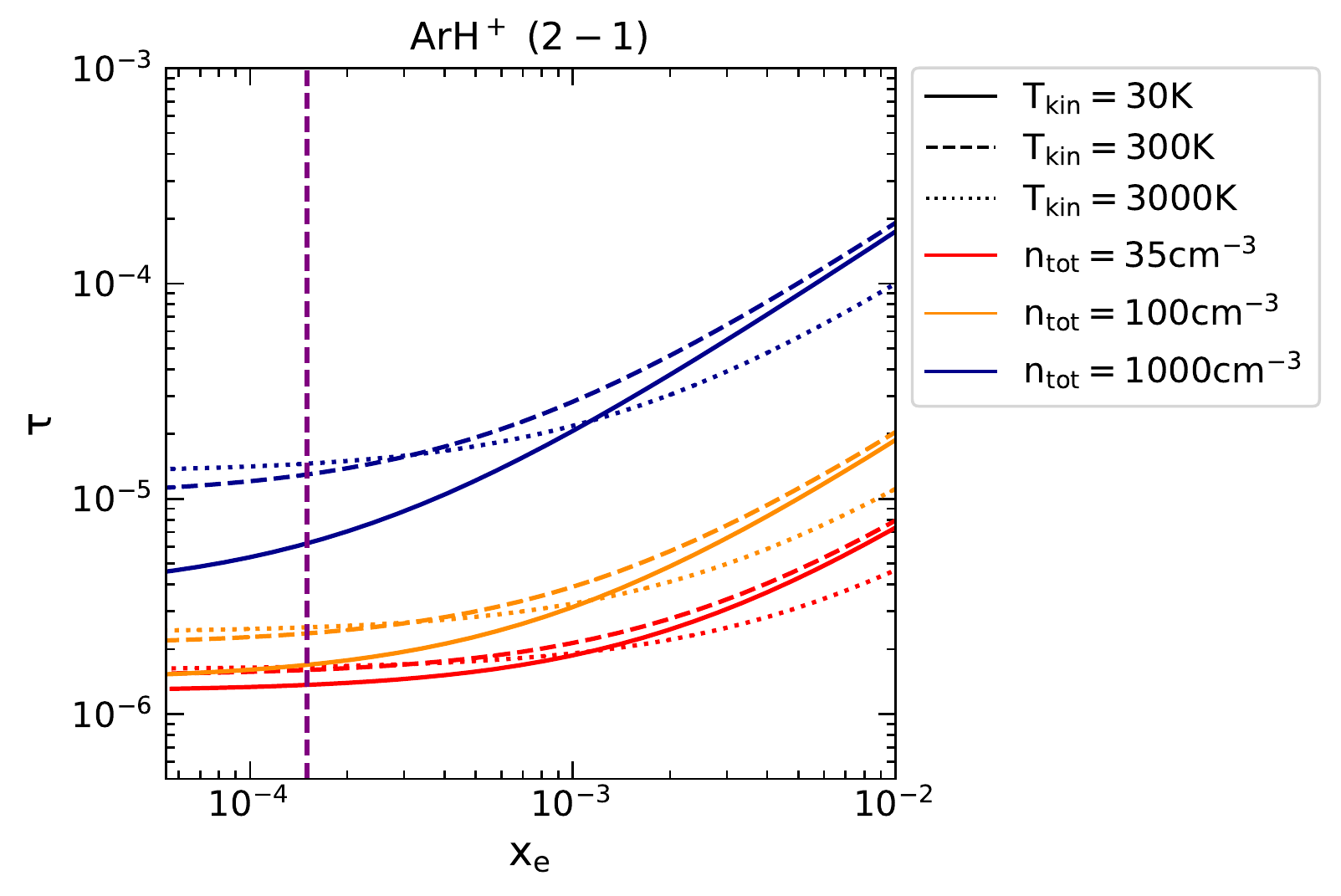}
    \caption{RADEX modelled excitation temperatures (left) and optical depths (right) as a function of electron density, for fixed values of the total gas density at 35 (red), 100 (orange) and 1000\,cm$^{-3}$ (dark blue) for the ArH$^+$ (2-1) transition. Furthermore, each model is run for fixed gas temperatures at 30 (solid), 300 (dashed) and 3000\,K (dotted), respectively. The horizontal dashed grey line marks an excitation temperature of 2.73\,K (in the left-hand panel) while the vertical dashed purple line indicates, $x_{\rm e} = 1.5 \times 10^{-4}$.}
    \label{fig:arhp_21_tau}
\end{figure*}

Using the same physical conditions as used to model the ArH$^+$ $1-0$ transition, we model the $2-1$ transition of ArH$^+$ near 1234.602\,GHz. The 
results are displayed in Fig.~\ref{fig:arhp_21_tau}. For all the models the excitation temperature for the $2-1$ line 
is higher, 
between ${\sim 8}$ and 18\,K, while the optical depths are extremely low, of the order of a few 10$^{-6}$ for 
total gas densities of 35\,cm$^{-3}$ and only an order of magnitude higher for models with total gas densities of 1000\,cm$^{-3}$. Therefore, as discussed by \citet{Dagdigian2018} given the low optical depths reproduced by the models, it is highly unlikely that the $2-1$ transition of ArH$^+$ can be observed in interstellar gas; so far it has only been detected (in emission) in the extreme circumnebular environment of the Crab \citep{barlow2013detection, Priestley2017}.

\subsection{\texorpdfstring{$\text{H}_{2}\text{O}^{+}$}{HtOp} ortho-to-para ratio}\label{subsec:h2o_analysis}
The H$_{2}$O$^+$ molecular ion exists in two symmetric states, o- and p-H$_{2}$O$^+$, that have opposing parities due to the interaction between the magnetic moment of the unpaired electron and protons. Studying the ratio of molecules in these two states which is reflected by the spin temperature, can provide insight into the formation pathway and thermodynamic properties of the gas. In the following paragraphs we determine the ortho-to-para ratio (OPR) of H$_{2}$O$^+$ and derive the nuclear spin temperature.

Unlike for H$_{2}$O, the lowest energy state of H$_{2}$O$^+$ corresponds to the lowest rotational energy level of its ortho state. This is a result of the molecule's $C_{2\varv}$ symmetry and $^{2}B_{1}$ ground state configuration. The fine structure levels of the o-H$_{2}$O$^+$ spin state, with a nuclear spin, $I$, of 1, further undergo HFS splitting while, p-H$_{2}$O$^+$ (with $I = 0$), does not. The lowest rotational energy level of the ortho spin state, ${N_{K_{a}K_{c}} = 1_{1,1}-0_{0,0}}$,  
lies at an energy level which is 30.1\,K lower energy than that of the lowest p-H$_{2}$O$^+$ level, 
$1_{1,0}-1_{0,1}$. 
Under the assumption that the rotational temperature is close to $T_{\text{CMB}} = 2.73\,$K, one would expect that most of the ions occupy either one of the H$_{2}$O$^+$ ground state levels. Similar to H$_{2}$ and H$_{2}$O, H$_{2}$O$^+$ is expected to exhibit an OPR of at least 3:1 \citep{townes1975microwave}. Typically, the conversion between the ortho and para states of H$_{2}$O$^+$ occurs via gas phase reactions with atomic hydrogen, as follows: 
\begin{equation}
    \text{p-H}_{2}\text{O}^+ + \text{H} \rightleftharpoons \text{o-H}_{2}\text{O}^+ + \text{H}
    \label{eqn:formation}
\end{equation}
while H$_{2}$O$^+$ energetically reacts with molecular hydrogen to yield H$_{3}$O$^+$. 

\noindent 
Moderately coupled via collisions, the OPR between the two spin states of H$_{2}$O$^+$ is given by,
\begin{equation}
    \text{OPR} \equiv \frac{Q_{\text{ortho}}}{Q_{\text{para}}}\text{exp}\left( -\Delta E / T_{\text{ns}}\right) \, ,
    \label{eqn:OPR}
\end{equation}
where $Q_{\text{ortho}}$ and $Q_{\text{para}}$ represent the partition functions of the respective spin isomers, $\Delta E$ is the energy difference between them, $\Delta E = -30.1~$K and $T_{\text{ns}}$ is the nuclear spin temperature. The value of $\Delta E$ is expressed as a negative quantity because, as discussed above, the lowest ortho state has a lower energy than the lowest para state. 
As discussed in Appendix A of \citet{schilke2010}, at low temperatures, the partition functions of the two states are governed by the degeneracy of their lowest fine-structure and HFS levels. Having the same quantum numbers and upper level degeneracies, the ratio of the partition functions approaches unity as the rotational temperature tends to 0~K.

\begin{figure*}
\sidecaption 
    \includegraphics[width=12cm]{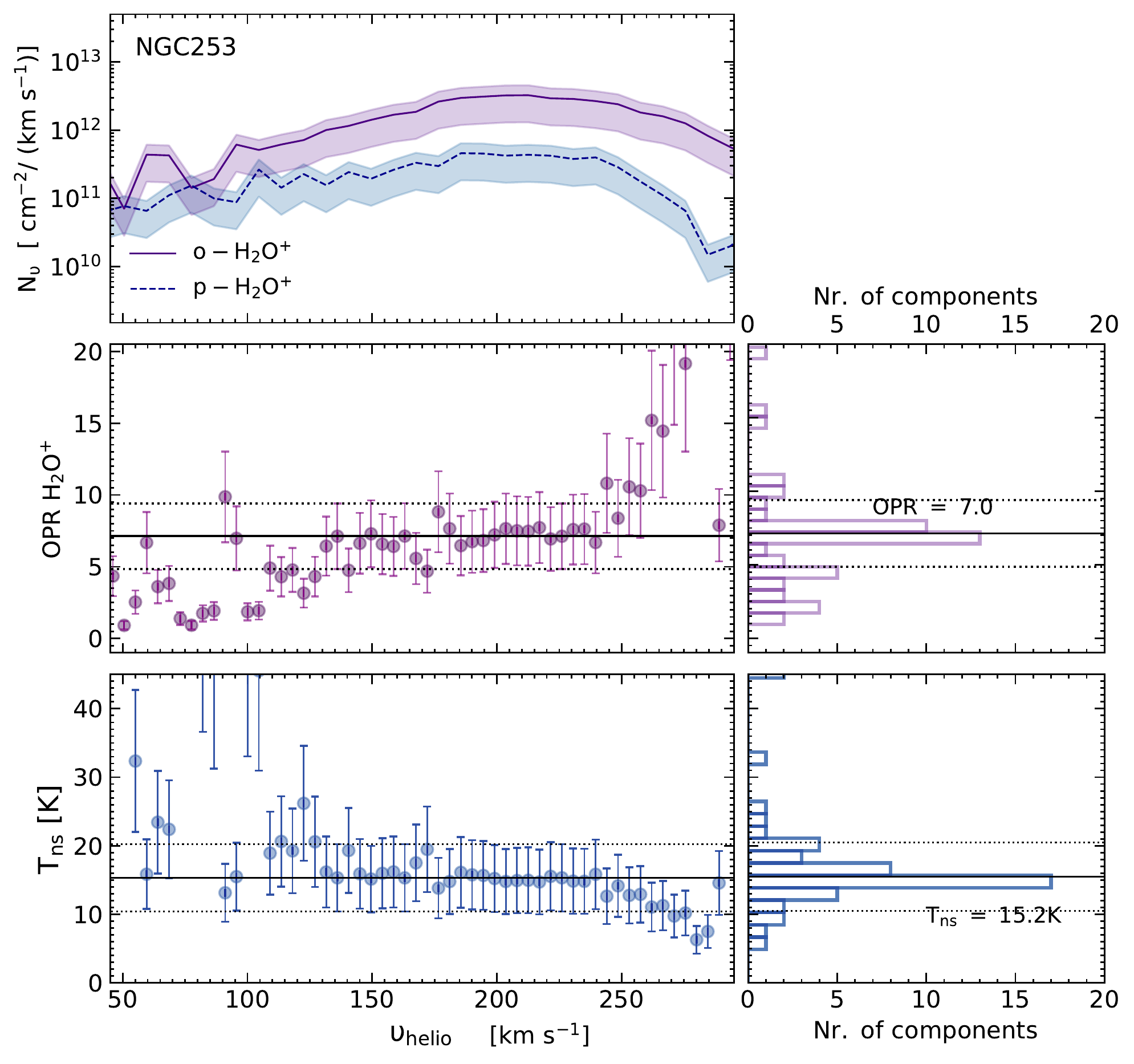}
    \caption{Top panel: Column density per velocity interval distribution of \ohtop (solid purple) and \phtop (dashed blue) over the entire velocity interval covering absorption towards NGC~253. Middle panel: OPR H$_{2}$O$^+$ distribution (left) and corresponding histogram (right). Bottom panel: Nuclear spin temperature distribution (left) and corresponding histogram (right). The median and $1~\sigma$ levels of the OPR and nuclear spin temperature are marked by solid and dotted black lines, respectively.}
    \label{fig:OPR_NGC253}
\end{figure*}

\begin{figure*}
    \sidecaption 
    \includegraphics[width=12cm]{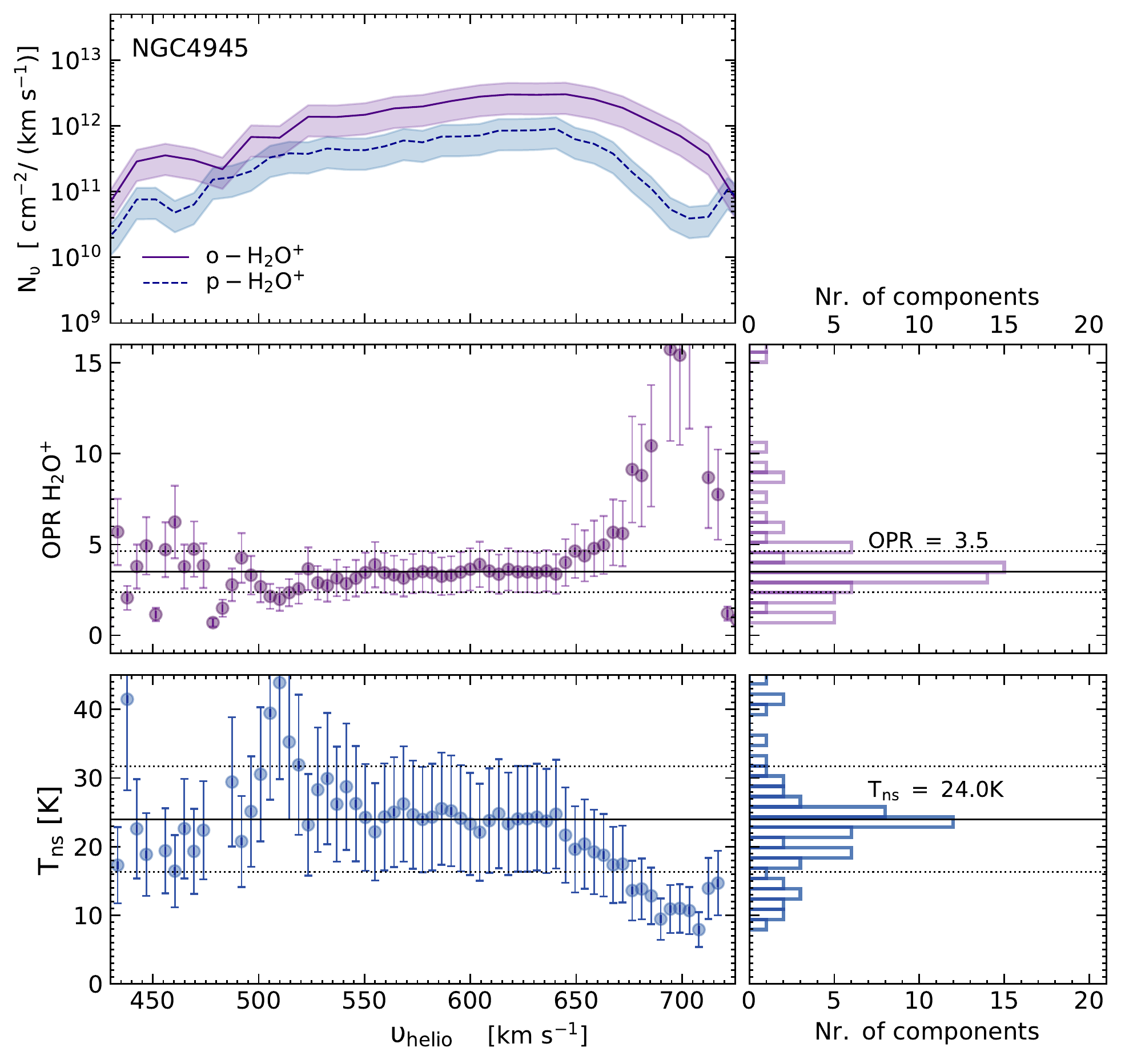}
    \caption{Same as Fig.~\ref{fig:OPR_NGC253}, but towards NGC~4945. The median value for $T_{\rm ns}$ derived from the distribution is the same as that obtained from deriving $T_{\rm ns}$ using the median OPR for H$_{2}$O$^+$.}
    \label{fig:OPR_NGC4945}
\end{figure*}

Figures~\ref{fig:OPR_NGC253} and \ref{fig:OPR_NGC4945} display the distribution of the derived column densities per velocity intervals (corresponding to the spacing of the velocity channel bins) for o- and p-H$_{2}$O$^{+}$, the OPR and the nuclear spin temperature. 
The OPRs determined for NGC~253 are quite large, with a mean value of roughly 7.0$\pm 3.1$, 
whereas, the mean value derived towards NGC~4945 is 3.5$\pm 1.2$. The derived OPRs are consistent with the equilibrium value of three within the quoted error bars and correspond to nuclear spin temperatures between 15 and 24~K. 

The very large values found for NGC~253 for LSR velocities ${>\!150}$~km~s$^{-1}$ can be partly attributed to the uncertainties resulting from modelling the emission component observed in the P-Cygni profiles of its o- and p-H$_{2}$O$^+$ spectra. 
Theoretical studies carried out by \citet{Tanaka2013} on the ortho-to-para exchange reactions of H$_{2}$O$^+$ reveal quite low reaction rates, which implies that the OPR of H$_{2}$O$^+$ does not significantly deviate from the equilibrium value of three, particularly at low temperatures, lower than the values we find for NGC 253. 
In diffuse regions affected by X-rays and/or cosmic-rays, the chemistry leading to H$_{2}$O$^+$ is initiated by the charge transfer between ionised hydrogen and oxygen atoms to form O$^+$, which is subsequently followed by reactions with  H$_{2}$ to first form OH$^+$ and then H$_{2}$O$^+$. Alternatively, in regions with higher molecular fractions, both OH$^+$ as well as H$_{2}$O$^+$ can be formed via reactions between H$_{3}^+$ and oxygen atoms \citep[][and references therein]{Hollenbach2012}. Lastly, H$_{3}$O$^+$ can also be formed from hydrogen abstraction reactions of H$_{2}$O$^+$ in denser environments. If the exchange reaction is faster than other competing reactions then the observed upper limits of the OPR, point to gas kinetic temperatures that are greater than $\sim$20~K. 

Since the diffuse and dense gas volumes along the sight lines towards the galaxies under study here are well mixed, the OPR of parent species such as H$_{3}^+$, H$_{2}$, and H$_{2}$O, from which H$_{2}$O$^+$ originates, may impact the observed OPR of H$_{2}$O$^+$ \citealp[see for example the studies on the OPR of H$_{2}$ presented by][]{flower2006importance}. Destruction reactions of both the ortho- and para- forms of H$_{3}^+$ via dissociative recombination have been experimentally investigated by \citet{glosik2010binary}, who observe a preferential recombination of this molecule's para-state compared to that of its ortho-state, which can also lead to OPRs \textgreater 3. Recently, \citet{Novotny2019} found a similar dependence on the rotational state, for the dissociative recombination of HeH$^+$ ions. However, a better understanding of the OPR of H$_{2}$O$^+$ requires a detailed understanding (not only) of what fraction of the observed H$_{2}$O$^+$ is formed along each of the different chemical pathways towards its formation, alongside the efficiency with which either H$_{2}$O$^+$ state (ortho or para) is formed in both quiescent and turbulent and/or shocked regions, but also the destruction of this ionic species which occurs primarily via dissociative recombination and hydrogen abstraction reactions with H$_{2}$. In Appendix~\ref{appendix:h2op_steadystate_chem} we briefly evaluate deviations in the observed OPR of H$_{2}$O$^+$ from the original OPR at formation, by carrying out a steady-state analysis of H$_{2}$O$^+$ chemistry.

\begin{figure}[ht!]
    \centering
    \includegraphics[width= 0.45\textwidth]{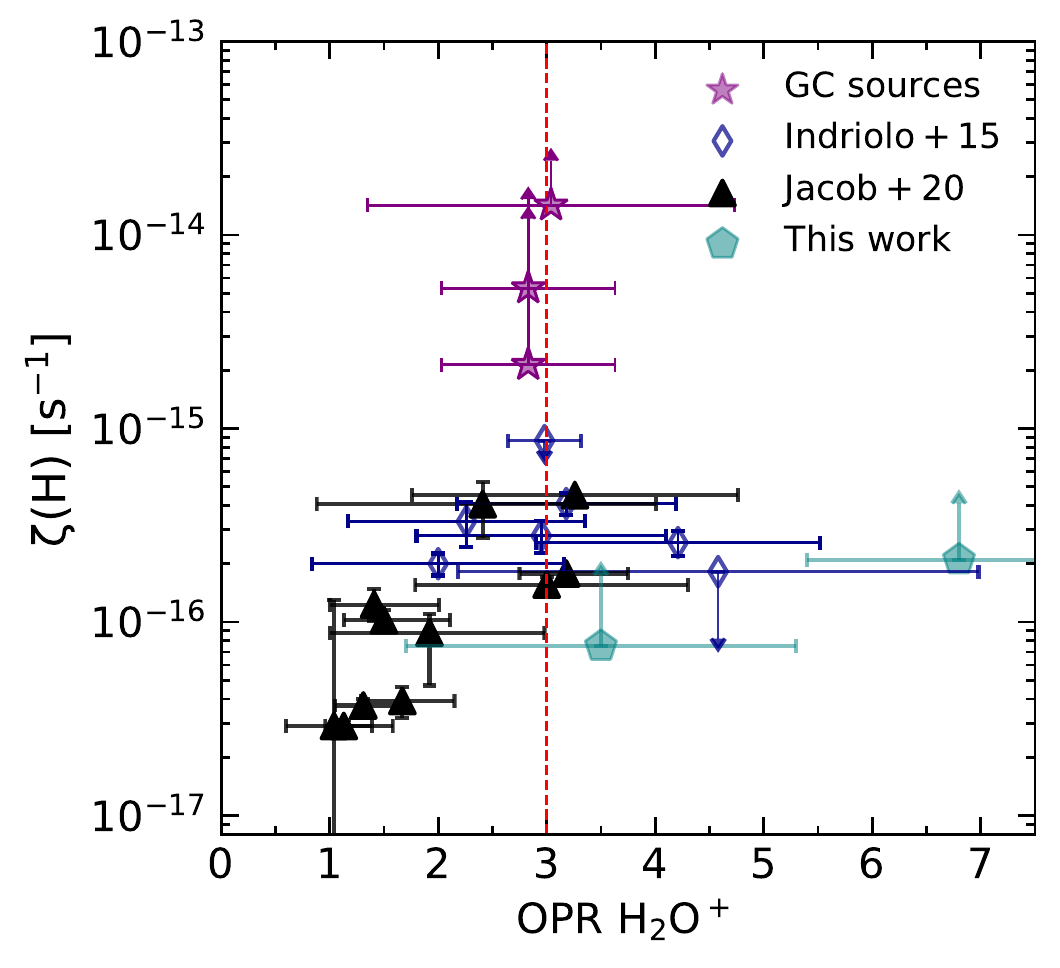}
    \caption{Observed OPR of H$_{2}$O$^+$ versus the cosmic-ray ionisation rates derived using the steady-state analysis of the OH$^+$ and H$_{2}$O$^+$. The results from this work are marked by teal pentagons while the blue diamonds and black triangles display the results for sight lines studied in the Milky Way by \citet{indriolo2015herschel} and \citet{Jacob2020Arhp}. For comparison we display the OPR of H$_{2}$O$^+$ also derived by \citet{indriolo2015herschel} towards Galactic centre (GC) sight lines using purple stars. The dashed red line marks the equilibrium OPR value of three.
    }
    \label{fig:CRIR_OPR}
\end{figure}

Furthermore, investigating the influence of cosmic-ray ionisation on the OPR of H$_{2}$O$^+$ for star-forming regions within the Milky Way, \citet{Jacob2020Arhp} find that the OPR of H$_{2}$O$^+$ thermalises to the equilibrium value of three with increasing rates of cosmic-ray ionisation up to a value of ${\sim\!2 \times 10^{-16}\,}$s$^{-1}$, beyond which there exists no credible correlation. The correlation between cosmic-ray ionisation rates and OPRs suggests that the higher abundances of atomic hydrogen in regions exposed to higher cosmic-ray fluxes are able to efficiently drive the proton-exchange reaction causing a change in the OPR. Putting the current results into context, we find the data points for both studied galaxies to lie in the thermalised region of Fig.~\ref{fig:CRIR_OPR}, with OPRs near three within the uncertainties. Given the uncertainties in the column densities derived from both the H{\small I} and the OH$^+$ line profiles, which are used to compute the cosmic-ray ionisation rates, we represent the cosmic-ray ionisation rates as lower limits. In analogy to the Milky Way, we would expect the cosmic-ray ionisation rates to be higher towards the centres of these galaxies, however in contrast, we derive values closer to the canonical values derived towards the disk of the Milky Way \citep{indriolo2015herschel, Jacob2020Arhp}, outside of the central molecular zone. 

\section{Discussion} \label{sec:discussion}

The gas properties of the ArH$^+$ bearing cloud volumes that we have derived in Sect.~\ref{subsec:gas_properties} towards the nearby galaxies studied in this work are comparable to those found by \citet{schilke2014ubiquitous} and \citet{Jacob2020Arhp} for clouds along the LOS towards star forming regions in the disk of the Milky Way. Furthermore, the cosmic-ray ionisation rates derived towards both NGC~253 and NGC~4945 in Sect.~\ref{subsec:CRIR} are comparable to the average cosmic-ray ionisation rate derived towards the disk of the Milky Way, of ($2.3\pm0.3$)$\times10^{-16}$~s$^{-1}$ \citep{indriolo2015herschel, Jacob2020Arhp}. The molecular gas traced by HF has an almost two orders of magnitude higher column density than H{\small I} (as illustrated in Fig.~\ref{fig:coldens_distributions}), suggesting that most of the gas, by volume, along the LOS towards these galaxies is dense. Ostensibly, this brings into question whether ArH$^+$ is a probe of diffuse atomic gas with a very small molecular fraction (see Sect.~\ref{subsec:gas_properties}). However, as previously discussed by \citet{van2016ionization}, who observed OH$^+$ and H$_{2}$O$^+$, other well-known tracers of (predominantly) atomic gas, towards both NGC~253 and NGC~4945, the presence of gas with 
low $f_{{\rm H}_{2}}$ may suggest an inhomogeneous ISM with local pockets of diffuse clouds mixed with denser 
material. Perhaps, the marginally lower cosmic-ray ionisation rates that we find for the galaxies studied here, in comparison to the Milky Way, may be a result of the larger volumes of dense gas present along the LOS. We note, however, that values for the cosmic-ray ionisation rate estimated for the nuclear regions of ULIRGs like NGC~4418 and Mrk~231, which harbour much more extreme star bursts than NGC~253 and NGC~4945, are almost three orders of magnitude higher with a lower-limit of $\zeta_{\rm p}(\text{H})\!>\!10^{-13}\,$s$^{-1}$ \citep{Gonzalez2013,Gonzales2018}, similar to values derived towards the Milky Way's Galactic centre region, indicative of higher levels of energetic phenomena present at the centres of these galaxies in comparison to their disks. This is also the case towards the diffuse atomic gas rich lines-of-sight studied towards PKS~1830$-$211, who derived column densities for atomic gas tracers such as ArH$^+$ \citep{Mueller2015}, p-H$_{2}$O$^+$, and OH$^+$ \citep{Mueller2016} that are roughly an order of magnitude greater than those that we derived towards NGC~253 and NGC~4945 while the column density values for molecular gas tracers such as HF are comparable between both studies \citep[see Table~2 of ][]{Mueller2016}. This results in cosmic-ray ionisation rates of $\sim 2\times10^{-14}~{\rm s}^{-1}$ and s$\sim 3\times 10^{-15}~{\rm s}^{-1}$ for the SW and NE lines-of-sight, respectively.

In addition, while we quote LOS averaged values for the cosmic-ray ionisation rates towards both NGC~253 and NGC~4945 using the steady-state analysis of OH$^+$ and H$_{2}$O$^+$, their interpretation is not trivial as the cosmic-ray ionisation rate is likely to vary spatially, across these galaxies. This variation is best demonstrated by observations towards different molecular cloud sight lines within the Milky Way by \citet{indriolo2015herschel} and \citet{Jacob2020Arhp} and reflects the proximity of the studied molecular clouds to the nearest supernova remnants, which are likely sites for particle acceleration amongst other propagation effects. Therefore we expect the cosmic-ray ionisation rates determined from compact regions of star-formation like the central molecular zones of the galaxies studied here, to be orders of magnitude greater. In general with star-formation rates higher than that of the Milky Way by almost two orders of magnitude, one would anticipate higher cosmic-ray ionisation rates towards both NGC~253 and NGC~4945 since the injection of cosmic-rays is governed by star-formation and accelerated by associated shocks like supernova remnants and stellar wind bubbles.  Observationally, higher rates of cosmic-ray ionisation have been estimated towards the centre of NGC~253 through the detection of gamma-rays at TeV energies using ground based Cherenkov telescopes \citep{Acero2009}. Exhibiting both starburst and AGN activities the nuclear environment of NGC~4945 is more complex, as a significant portion of the injected cosmic-ray energy is used in the production of pions and $\gamma$-rays. \citet{Wojaczynski2017A} have shown that Seyfert galaxies like NGC~4945 have larger $\gamma$-ray luminosity's than the calorimetric limit which sets the total energy of cosmic-rays used for the production of pions assuming that each supernova explosion injects 10$^{50}$~ergs of cosmic-ray energy while that of starbursts like NGC~253 is 50~\% lower.

Moreover, while ionisation by cosmic-rays plays an important role in heating the gas in the nuclear environment of galaxies \citep{Bradford2003}, heating by X-rays \citep{Usero2004} and by stellar UV radiation \citep[in widespread photodissociation regions (PDRs)][]{Hollenbach1997}, dynamical shock heating \citep{Burillo2001} to a name a few, can all be prominent sources of heating in these regions and it is not possible to discriminate between their contributions merely by an analysis such as ours.

\section{Conclusions} \label{sec:conclusions}
Well established as a tracer for purely atomic gas, outside of the Milky Way the noble gas species ArH$^+$ had only been detected towards the intermediate redshift ${z = 0.89}$ gravitational lens system PKS 1830$-$211. In this work we extend the notion that ArH$^+$ is an ubiquitous tracer of diffuse atomic gas towards external galaxies by presenting observations of its ${J\!=\!1-0}$ transition.

 We present the detection of ArH$^+$ in absorption towards two nearby galaxies, NGC~253 and NGC~4945 and the non-detection towards Arp~220. In addition to ArH$^+$ we also report the detection of the ${J=3/2-1/2}$ transition of p-H$_{2}$O$^+$ at 607\,GHz towards the former. We compare the observed profiles of ArH$^+$ and p-H$_{2}$O$^+$ lines with those of the  H{\small I} 21\,cm line and lines from the well known atomic gas tracers OH$^+$ and o-H$_{2}$O$^+$ and the molecular gas tracer, HF. Using the cosmic-ray ionisation rates derived by analysing the steady-state chemistry of OH$^+$ and H$_{2}$O$^+$ and by assuming that the cloud populations bearing ArH$^+$ are exposed to the same cosmic-ray ionisation rates, we derive the molecular fraction of the gas traced by ArH$^+$ towards both galaxies to be between ${\!10^{-3}}$ and ${\sim\!10^{-2}}$. This is consistent with estimates made within the Milky Way by \citet{schilke2014ubiquitous, neufeld2016chemistry, Jacob2020Arhp}. From the detection of p-H$_{2}$O$^+$ we estimate the H$_{2}$O$^+$ OPR of 7.0 and 3.5 towards NGC~253 and NGC~4945, and subsequently derive values for the nuclear-spin temperature that are greater than 15 and 24\,K, respectively. The variation in the H$_{2}$O$^+$ OPR with cosmic-ray ionisation rates follows the trend displayed by sight line components towards star-forming regions within the Milky Way other than those towards the Galactic centre. This is likely because our observations are not sensitive to spatial variations along the sight lines studied and only probe the average properties.

In order to comment on the ubiquity and chemistry of ArH$^+$ and its possible role in probing the energetics of extragalactic sources, the searches for ArH$^+$ need to be extended to a wider source sample. Since its chemistry and properties are widely studied, alongside that of OH$^+$ and o-H$_{2}$O$^+$, it may be promising to search for ArH$^+$ towards a range of extragalactic sources hosting various levels of nuclear activity, which was a selection criterion for  the Herschel EXtraGALactic (HEXGAL) survey which was aimed at studying the physical and chemical composition of the ISM in galactic nuclei using HIFI spectroscopy.

\begin{acknowledgements}

This publication is based on data acquired with the Atacama Pathfinder Experiment (APEX) under the project id M9519C\_103. APEX is a collaboration between the Max-Planck-Institut f\"{u}r Radioastronomie, the European Southern Observatory, and the Onsala Space Observatory. We would like to express our gratitude to the APEX staff and science team for their continued assistance in carrying out the observations presented in this work. We would like to thank Dr.~Paul Dagdigian for providing us with the ArH$^+$ collisional rate coefficients and the anonymous referee for their careful review of the article and valuable input. We are thankful to the developers of the C++ and Python libraries and for  making them available as open-source software. In particular, this research has made use of the NumPy \citep{numpy}, SciPy \citep{scipy} and matplotlib \citep{matplotlib} packages.
\end{acknowledgements}
\bibliographystyle{aa} 
\bibliography{ref}

\begin{appendix}
\section{H{\small I} analysis}\label{appendix:hi_analysis}
\begin{figure}
    \centering
    \includegraphics[width=0.48\textwidth]{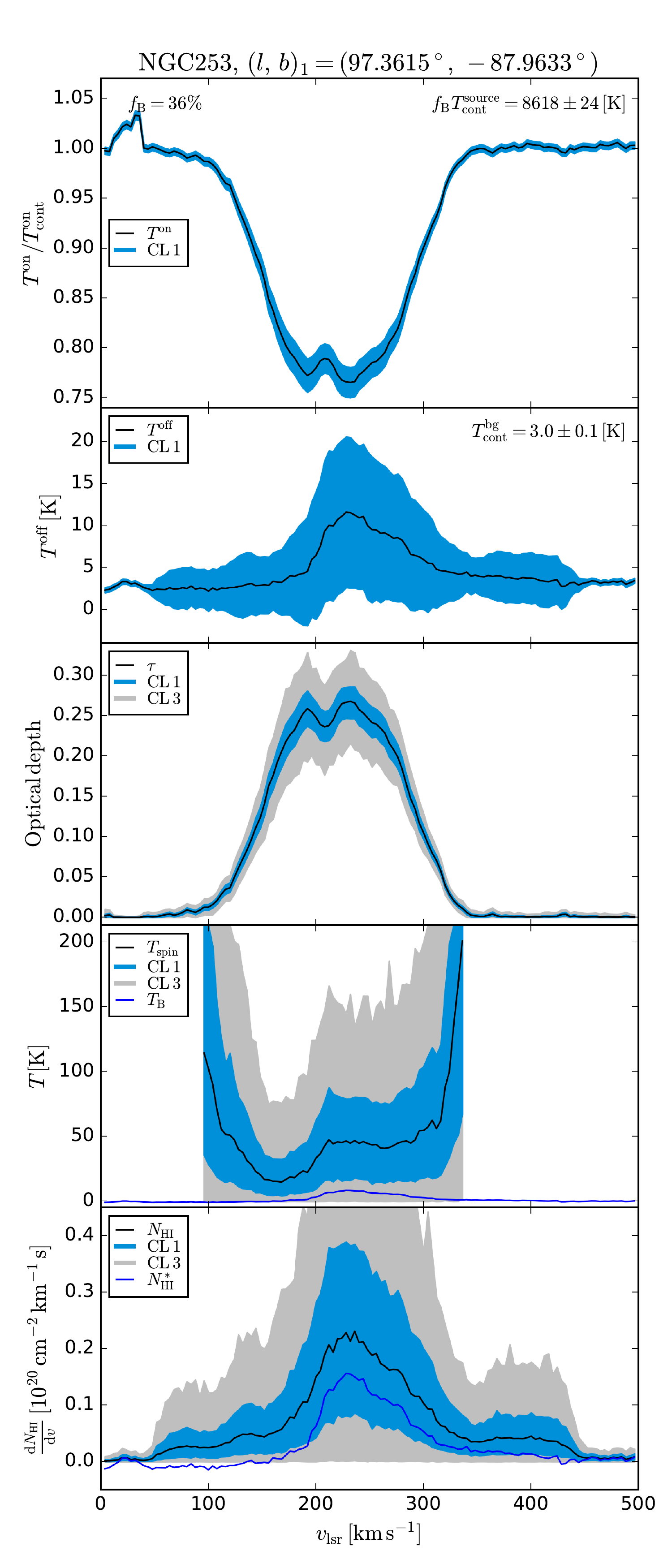}
    \caption{From top to bottom: H{\small I} absorption and emission spectra, optical depth, spin temperature, and H{\small I} column density towards NGC~253.
    }
    \label{fig:NGC253_HI}
\end{figure}

\begin{figure}
    \centering
    \includegraphics[width=0.48\textwidth]{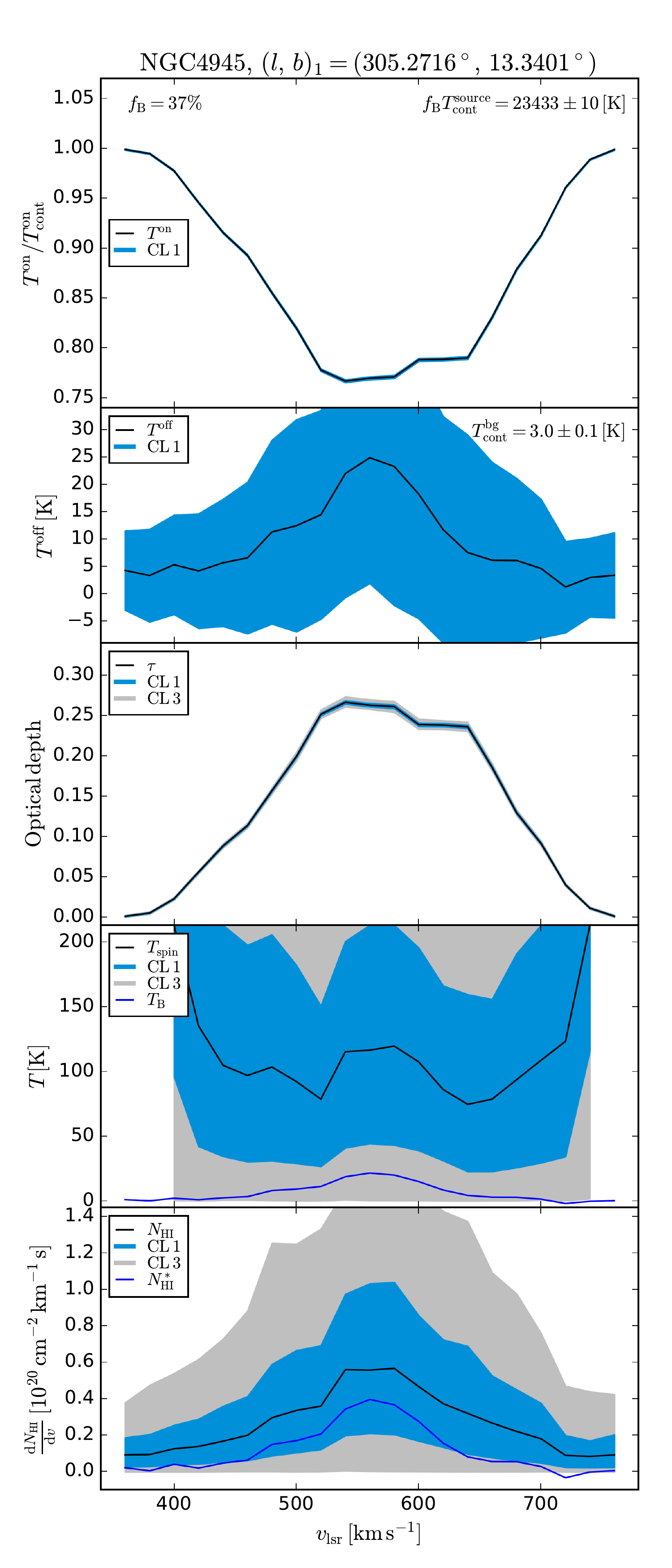}
    \caption{Same as Fig.~\ref{fig:NGC253_HI}, but towards NGC~4945. 
    }
    \label{fig:NGC4945_HI}
\end{figure}

In this appendix we discuss 
the H{\small I} 21\,cm line analysis. The H{\small{I}} absorption ($T_\text{on}$) and emission spectra ($T_\text{off}$), along with derived quantities such as the optical depth, spin temperature and column density are presented in Figs.~\ref{fig:NGC253_HI} and \ref{fig:NGC4945_HI} for NGC~253 and NGC~4945, respectively. The solid black curve in the top panel of these figures, represents the continuum normalised on-source absorption profiles ($T^{\text{on,obs}}/ T^{\text{sou,obs}}_{{\text{cont}}}$) smoothed to the resolution of the emission line data. The off-source spectrum is taken from a position next to the source of interest (well outside of the beam width). Similar to the Milky Way disk, the H{\small I} data towards these galaxies are subject to strong spatial fluctuations and using a single off position would be error prone. Therefore, following \citet{winkel2017hydrogen}, a spatial filtering technique using a ring (or doughnut-shaped) kernel was applied to obtain an interpolated brightness temperature, as shown in the second panel (black curve). For comparison, in the panel displaying spin temperatures (computed for regions with $\tau >0.01$), we also plot the brightness temperature in blue and lastly, the column density panel also displays the uncorrected column density profile, $N^{*}_{\text{HI}}$. The blue and grey shaded regions indicate the $1\sigma$ (68\% percentile) and $3\sigma$ (99.7\% percentile) confidence intervals, respectively. A detailed description is presented in \citet{winkel2017hydrogen}. As a caveat in this analysis, we mention here that this filtering technique may not be what is best suited to describe the often observed, `butterfly'-shaped distribution of gas in these external galaxies and a proper treatment of the same, would require the use of a tilted ring model. In the on-source H{\small I} absorption profile towards NGC~253, we note the appearance of an emission feature between 5 and 40\,km\,s$^{-1}$, which is merely an artefact arising from the post-processing (cleaning) of the data or is caused by H{\small I} emission from the disk of the galaxy.


\section{Continuum level uncertainties}\label{sec:continuum_level_incertainties}
\begin{figure}
    \centering
    \includegraphics[width=0.45\textwidth]{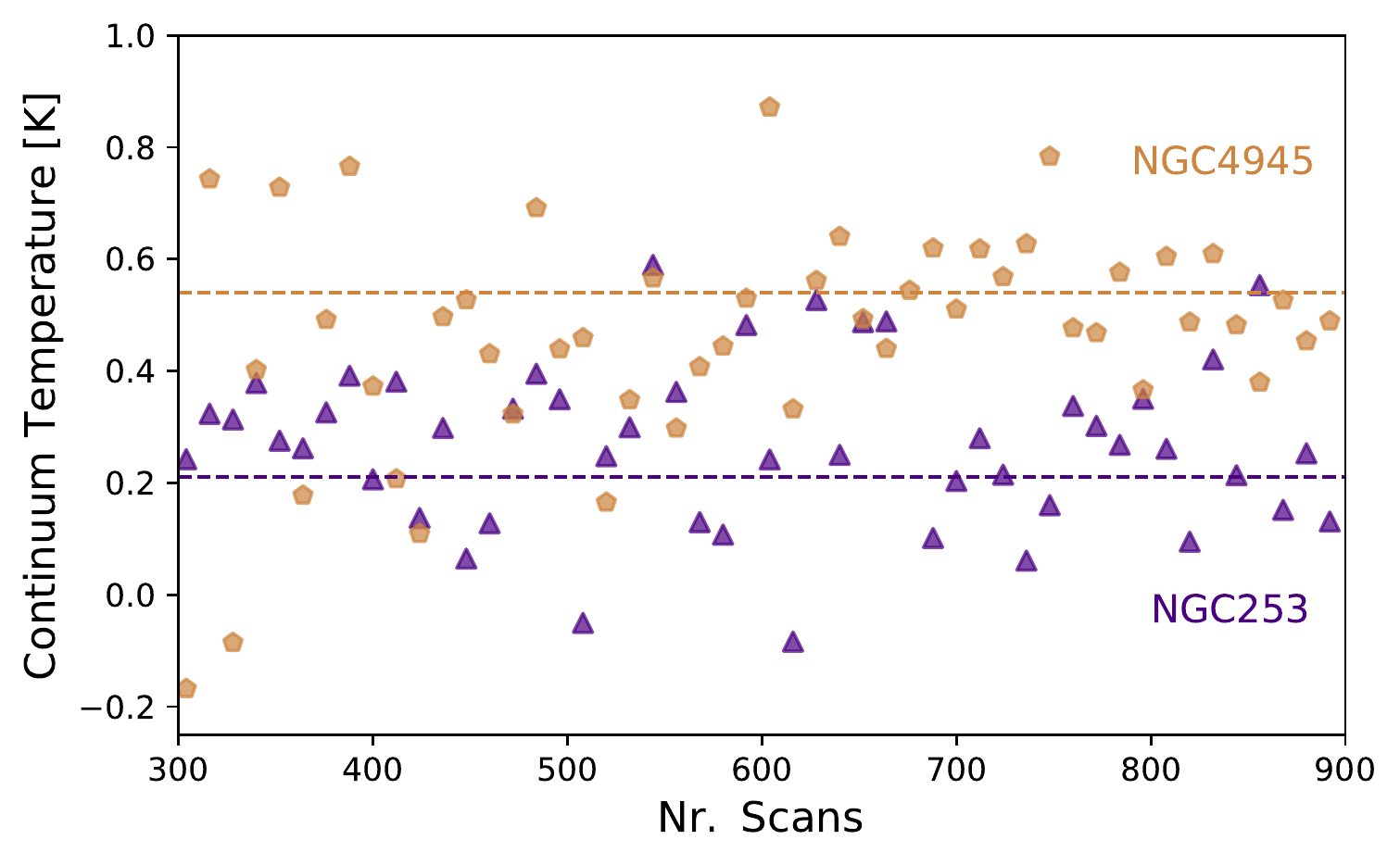}
    \caption{Continuum fluctuations across scans for NGC~253 (purple triangles) and NGC~4945 (yellow pentagons) with the dashed lines indicating the median continuum levels for each source.}
    \label{fig:continuum_scans_scatter}
\end{figure}

We now briefly assess the reliability of the absolute calibration of the continuum brightness temperatures used in our analysis that were derived from the sub-mm spectra of the ArH$^+$ and p-H$_2$O$^+$ lines taken with the APEX telescope. In Fig.~\ref{fig:continuum_scans_scatter}, for both NGC~253 and NGC~4945, we plot the continuum brightness temperatures measured for all individual scans versus the time over velocity intervals used to determine the continuum level and baseline. The scatter across the different scans displayed is found to lie between 11 and 17\% for the two sources with an average value of 14\%, thereby yielding a fairly stable continuum level.

\section{Steady-state chemistry of \texorpdfstring{ H$_{2}$O$^+$}{h2op}}\label{appendix:h2op_steadystate_chem}

As discussed in \citet{Gerin2013}, the relationship between the initial OPR of the H$_{2}$O$^+$ ions when formed, OPR$_{0}$, with that which is observed, depends upon the relative rates of these processes. We adapt the steady-state analysis of H$_{2}$Cl$^+$ presented in \citet{Neufeld2015} but for H$_{2}$O$^+$, to estimate the relative importance of the different reactions in thermalising the OPR. Hence, the OPR of H$_{2}$O$^+$ is given by,

\begin{equation}
    \text{OPR} = \frac{k_{\rm po}n(\text H) + k_{\text dr}n_{e}\text{OPR}_{0} + k_{\text ha}n(\rm H_{2})OPR_{0}}{k_{\rm op}n(\text H)+ k_{\text dr}n_{e} + k_{\text ha}n(\rm H_{2})} \, ,
\end{equation}
where $k_{\rm po}$ and $k_{\rm op}$ are the rate coefficients for the forward and backward reactions in (\ref{eqn:formation}) and $k_{\rm dr}$ and $k_{\rm ha}$ are the reaction rates for the dissociative recombination and hydrogen abstraction reactions, respectively. Assuming a detailed balance, such that ${k_{\rm po} = k_{\rm op}{\rm OPR}_{\rm LTE}({\rm H}_{2}{\rm O}^+)}$, the above equation can be re-written as
\begin{equation}
    \text{OPR} = x\text{OPR}_{\rm LTE} + (1-x) \text{OPR}_{0}\, ,
\end{equation}
where $x = k_{\text {op}}n(\text{H})/\left(k_{\text {op}}n(\text{H}) + k_{\text {dr}}n_e + k_{\text {ha}}n(\text{H}_{2})\right)$. The rate coefficient for the proton exchange reaction (\ref{eqn:formation}), $k_{\rm op}$, has not been measured but is assumed to be of the order of $10^{-10}$~cm$^{3}$~s$^{-1}$ in view of the rate coefficients determined for exchange reactions of other related species such as H$_{2}$, p-H$_2$ + H$^+$ $\rightleftharpoons$ o-H$_2$ + H$^+$, for which $k = 4.15\times10^{-10}$~cm$^3$~s$^{-1}$ \citep{Honvault2011}. While the rate coefficient for the dissociative recombination reaction is $k_{\rm dr} \sim 7.44\times10^{-7}$~cm$^3$~s$^{-1}$, that for the abstraction reaction is $k_{\rm ha} = 6.1\times10^{-10}$~cm$^3$~s$^{-1}$ (computed at a gas temperature of 100~K, typical for diffuse gas environments), competes with $k_{\rm op}$. Using values for $n_{\rm H}$, $n_{\rm H_{2}}$, and $n_{e}$ as discussed in Sects.~\ref{subsec:CRIR} and \ref{subsec:gas_properties} and $k_{\rm op} = 4.15\times 10^{-10}$~cm~s$^{-3}$, we estimate $x$ to be 0.74. This implies that, for values of $k_{\rm op} > 4.15\times 10^{-10}$~cm~s$^{-3}$, $x \rightarrow 1$ and that the OPR is close to the values at LTE. On the contrary, if $k_{\rm op} > 4.15\times 10^{-10}$~cm~s$^{-3}$ then $x \rightarrow 0$ implying that the OPR is close to OPR$_{0}$. Given the many uncertainties involved in this calculation we can only conclude that the OPR derived using Eq.~\ref{eqn:OPR} merely represents a limit.

\end{appendix}

\end{document}